# Elliptical model of cutoff boundaries for the solar energetic particles measured by POES satellites in December 2006

A. V. Dmitriev, P. T. Jayachandran, L.-C. Tsai

```
A set of FORTRAN subroutines to calculate the cutoff latitude
for charged particle penetration
into the polar cap of the Earth's magnetosphere.
Dmitriev, A. V., P. T. Jayachandran, and L.-C. Tsai (2010), Elliptical model of
cutoff boundaries for the solar energetic particles measured by POES satellites
in December 2006, J. Geophys. Res., 115, A12244, doi:10.1029/2010JA015380.

Written by Alexei Dmitriev <alexei_dmitriev@yahoo.com>

PCNP***************************************************
PCNP(cLat,R,rMLT,Dst,vKp,PS)
proton cutoff invariant colatitude cLat in the Northern polar cap
proton energy range: from 0.03 to 100 MeV

      subroutine PCNP
      o( cLat,     ! cutoff invariant co-latitude (deg)
      i  R,  ! rigidity in MV R=sqrt(E*E+2*988.*E), E in MeV
      i  rMLT,     ! rotated magnetic local time (deg): rMLT=(MLT-6)*15. (deg)
      i  Dst,vKp,! geomagnetic indices Dst and digital Kp (Kp*10=0~90)
      !  PS    )! geodipole tilt angle (deg)

      radeg=180./acos(-1.)    ! radians to degrees
      gR=alog(R)         ! natural logarithm exp(alog(R))=R

semiaxes
magnetic quiet
      as=6.2338E-005*R +0.0240963 -exp(-0.0219198*R)*0.0126371
      as=asin(as**0.25)*radeg
      bs=6.20775E-005*R +0.0211002 -exp(-0.018097*R)*0.0118746
      bs=asin(bs**0.25)*radeg

magnetic disturbances
as
      asDst=-9.738E-02 +1.156E-02*gR
      asKp=8.550E-02 -7.461E-03*gR
      das=-0.5683+asDst*Dst+asKp*vKp
      as=as+das
bs
      bsDst=-9.539E-02 +1.155E-02*gR
      bsKp=7.657E-02 -7.487E-03*gR
      dbs=-0.6483+bsDst*Dst+bsKp*vKp
      bs=bs+dbs
Xo
      XoA=-2.907E-01  +1.293E-01*gR
      XoDst=3.685E-02 -5.437E-03*gR
      XoKp=-2.936E-02 +3.400E-03*gR
      Xo=XoA+XoDst*Dst+XoKp*vKp
Yo
      YoA=-9.265E+00 + 1.526E+00*gR
      YoDst=-5.430E-02 + 8.268E-03*gR
      YoKp= -6.602E-02 + 1.048E-02*gR
      YoPS= -7.400E-02 + 1.115E-02*gR
      Yo=YoA + YoDst*Dst + YoKp*vKp + YoPS*PS
fi
```

```
        fi=-5.487E+01 + 1.134E+01*gR
ellipse
        a=rMLT/radeg
        fir=fi/radeg
        x=Xo + as*cos(a)*cos(fir) - bs*sin(a)*sin(fir)
        y=Yo + as*cos(a)*sin(fir) + bs*sin(a)*cos(fir)
        cLat=sqrt(x*x+y*y)

        return
        end

PCNP************************************************

PCSP************************************************
PCSP(cLat,R,rMLT,Dst,vKp,PS)
proton cutoff invariant colatitude cLat in the Southern polar cap
proton energy range: from 0.03 to 100 MeV

        subroutine PCSP
      o( cLat,      ! cutoff invariant co-latitude (deg)
      i  R,  ! rigidity in MV R=sqrt(E*E+2*988.*E)
      i  rMLT,      ! rotated magnetic local time (deg): rMLT=(MLT-6)*15.
      i  Dst,vKp,! geomagnetic indices Dst and digital Kp (Kp*10=0~90)
      !  PS    )! geodipole tilt angle (deg)

        radeg=180./acos(-1.)    ! radians to degrees
        gR=alog(R)          ! natural logarithm exp(alog(R))=R

semiaxes
magnetic quiet
        as=6.62736E-005*R +0.024596 -exp(-0.0127964*R)*0.00922368
        as=asin(as**0.25)*radeg
        bs=7.01822E-005*R +0.0194598 -exp(-0.00817293*R)*0.00895325
        bs=asin(bs**0.25)*radeg

magnetic disturbances
as
        asDst=-6.40E-02 +5.97E-03*gR
        asKp=  1.15E-01 -1.37E-02*gR
        das=-0.740429
        das=das + asDst*Dst + asKp*vKp
        as=as+das
bs
        bsDst=-5.139E-02 +4.454E-03*gR
        bsKp=  5.207E-02 -3.509E-03*gR
        dbs=  -0.4 +bsDst*Dst+bsKp*vKp
        bs=bs+dbs
Xo
        XoA= -2.299E-01 +3.132E-02*gR
        XoA= -0.302
        XoDst=5.526E-02 -7.292E-03*gR
        XoKp=-4.107E-02 +5.514E-03*gR
        Xo=XoA + XoDst*Dst + XoKp*vKp
Yo
        YoA=  -8.544E+00 + 1.432E+00*gR
        YoDst=-3.928E-02 + 4.325E-03*gR !best
        YoKp= -7.893E-02 + 1.172E-02*gR     !orig
        YoPS= -7.273E-02 + 1.295E-02*gR
        Yo=YoA + YoDst*Dst + YoKp*vKp + YoPS*PS
fi
```

```
        fi=-4.450E+01 + 7.830E+00*gR

ellipse
        a=rMLT/radeg
        fir=fi/radeg
        x=Xo + as*cos(a)*cos(fir) - bs*sin(a)*sin(fir)
        y=Yo + as*cos(a)*sin(fir) + bs*sin(a)*cos(fir)
        cLat=sqrt(x*x+y*y)

        return
        end

PCSP***************************************************

PCNE***************************************************
PCNe(cLat,R,rMLT,Dst,AE,vKp,PS)
electron cutoff invariant colatitude cLat in the Northern polar cap
electron energy range: from 0.03 to 1 MeV

        subroutine PCNe
      o( cLat,          ! cutoff invariant co-latitude (deg)
      i  R,      ! (not used) rigidity in MV R=sqrt(E*E+2*988.*E)
      i  rMLT,          ! rotated magnetic local time (deg): rMLT=(MLT-6)*15.
      i  Dst,AE,vKp, ! geomagnetic indices Dst, AE and digital Kp (Kp*10=0~90)
      !  PS          )! geodipole tilt angle (deg)

        radeg=180./acos(-1.)    ! radians to degrees

          as=1.89E+01 - 4.30E-02*Dst + 9.86E-02*vKp
        bs=1.71E+01 - 4.97E-02*Dst + 8.42E-02*vKp
        Xo= 7.58E-01 + 4.82E-04*Dst - 2.58E-02*vKp
        Yo=-5.47E+00 + 4.67E-04*AE  - 4.68E-02*PS
        fi= 1.76E+01 - 6.70E-01*Dst + 1.78E+00*PS
        fir=fi/radeg
          a=rMLT/radeg
        x=Xo + as*cos(a)*cos(fir) - bs*sin(a)*sin(fir)
        y=Yo + as*cos(a)*sin(fir) + bs*sin(a)*cos(fir)
        cLat=sqrt(x*x+y*y)

        return
        end

PCNE***************************************************

PCSE***************************************************
PCSe(cLat,R,rMLT,Dst,AE,vKp,PS)
electron cutoff invariant colatitude cLat in the Southern polar cap
electron energy range: from 0.03 to 1 MeV

        subroutine PCSe
      o( cLat,          ! cutoff invariant co-latitude (deg)
      i  R,      ! (not used) rigidity in MV R=sqrt(E*E+2*988.*E)
      i  rMLT,          ! rotated magnetic local time (deg): rMLT=(MLT-6)*15.
      i  Dst,AE,vKp, ! geomagnetic indices Dst, AE and digital Kp (Kp*10=0~90)
      !  PS          )! geodipole tilt angle (deg)

        radeg=180./acos(-1.)    ! radians to degrees

          as=1.847E+01 -2.826E-02*Dst +1.212E-01*vKp
        bs=1.695E+01 -1.885E-02*Dst +8.867E-02*vKp
        Xo=5.055E-01 +2.731E-02*Dst -2.545E-02*vKp
```

```
Yo=-5.595E+00+3.588E-04*AE  -4.02E-02*PS
fi=-3.124E+01+2.252E-01*Dst -7.88E-01*PS
fir=fi/radeg
  a=rMLT/radeg
x=Xo + as*cos(a)*cos(fir) - bs*sin(a)*sin(fir)
y=Yo + as*cos(a)*sin(fir) + bs*sin(a)*cos(fir)
cLat=sqrt(x*x+y*y)

return
end
```

PCSE**************************************************

**Elliptical model of cutoff boundaries for the solar energetic particles measured by POES satellites in December 2006**

A. V. Dmitriev[1,2], P. T. Jayachandran[3], L.-C. Tsai[4]

[1]*Institute of Space Sciences, National Central University, Chung-Li, Taiwan*

[2]*Skobeltsyn Institute of Nuclear Physics Moscow State University, Moscow, Russia*

[3]*Physics Department of University of New Brunswick, Canada*

[4]*Center for Space and Remote Sensing Research, National Central University, Chung-Li, Taiwan*

________

A. V. Dmitriev, Institute of Space Science National Central University, Chung-Li, 320, Taiwan, also D.V. Skobeltsyn Institute of Nuclear Physics, Moscow State University, Russia (e-mail: dalex@jupiter.ss.ncu.edu.tw)

P. T. Jayachandran, Physics Department of University of New Brunswick, Canada (e-mail: jaya@unb.ca)

L.-C. Tsai, Center for Space and Remote Sensing Research, National Central University, Chung-Li, Taiwan (e-mail: lctsai@csrsr.ncu.edu.tw)

**Abstract** Experimental data from a constellation of five NOAA POES satellites were used for studying the penetration of solar energetic particles (SEP) to high latitudes during long-lasting SEP events on December 5 to 15, 2006. We determined cutoff latitudes for electrons with energies >100 keV and > 300 keV, and protons with energies from 240 keV to >140 MeV. The large number of satellites allowed us to derive snap shots of the cutoff boundaries with 1-hour time resolution. The boundaries were fitted well by ellipses. Based on the elliptical approach, we developed a model of cutoff latitudes for protons and electrons in the northern and southern hemispheres. The cutoff latitude is represented as a function of rigidity, $R$, of particles, MLT, geomagnetic indices: $Dst$, $Kp$, $AE$, and dipole tilt angle $PS$. The model predicts tailward and duskward shifting of the cutoff boundaries in relation to intensification of the cross-tail current, field-aligned currents, and symmetrical and asymmetrical parts of the ring current. The model was applied for prediction of a polar cap absorption (PCA) effects observed at high latitudes by CADI network of ionosondes. It was found that the PCA effects are related mainly to intense fluxes of >2.5 MeV protons and >100 keV electrons, which contribute mostly to the ionization of ionospheric $D$-layer at altitudes of ~75 to 85 km. This finding was confirmed independently by FORMOSAT-3/COSMIC observations of the SEP-associated enhancements of electron content at altitudes of ~80 km.

*Keywords:* solar energetic particles, geomagnetic cutoff, polar cap absorption

# 1. Introduction

Solar energetic particles (SEP), arriving at the Earth, are able to reach the low altitudes and affect the ionosphere and atmosphere at high and middle latitudes [*e.g. Galand*, 2001]. The penetration is mainly controlled by the magnetospheric magnetic field [*e.g. Smart et al.*, 2000], and depends also on characteristics of the interplanetary magnetic field (IMF) and on pitch angular distribution of incident particles [*Blake et al.*, 2001]. The basic parameter of a charged particle motion in the magnetic field is a rigidity, defined as the particle momentum divided by its charge. For single-charge particles, the rigidity $R$ is related to kinetic energy $E$ as $R = \sqrt{E^2 + 2E_0 E}$, where $E_0$ is the rest mass of particle. If the energies $E$ and $E_0$ are expressed in MeV, the rigidity is expressed in MV.

All particles having the same rigidity, charge sign and initial conditions should have identical trajectories of adiabatic motion in the magnetic field. Penetration of charged particles from the interplanetary medium into the magnetosphere is restricted by cutoff latitude, the minimal latitude at which a particle with a given rigidity is able to reach the Earth's surface. For instance, relativistic protons with rigidity of ~14 GV can reach the Earth at the equator, i.e. their cutoff latitude is zero. Particles with lower rigidities (and energies) will cut off at higher latitudes.

The relationship between the rigidity and cutoff latitude is a subject of numerous theoretical and experimental studies. As a first approach, it can be expressed as $R = f(\lambda) + D$, where $f(\lambda)$ is a function of invariant latitude $\lambda$, and $D$ is a parameter depending on configuration of the geomagnetic field. The invariant latitude is related to a McIlwian $L$ parameter as $L = \cos^{-2}(\lambda)$. The $L$ parameter is calculated from a geomagnetic field model. In a dipole approach, which is mainly proper for the inner magnetosphere, *Smart and Shea* [1994] derived the following relationship:

$$R(\mathrm{GV}) = 14.5 \cos^4(\lambda) \quad (1).$$

The coefficient of 14.5 GV is proportional to the dipole moment, which decreases gradually with a rate of ~0.1% per year [*Shea and Smart*, 2004].

At middle and high latitudes ($\lambda > 50°$), the geomagnetic field is different from the dipole: field lines are compressed at the dayside by the solar wind impact and stretched toward the magnetotail on the nightside. Hence, the cutoff latitudes at the nightside should be located lower than those predicted from the pure dipole approach. Models of the external magnetic field, such as Tsyganenko models or global MHD magnetospheric codes, are used for simulations of the SEP penetration in the magnetosphere [*Blake et al.*, 2001; *Kahler and Ling*, 2002; *Kress et al.*, 2004; *Birch et al.*, 2005;

*Smart et al.*, 2006]. As a rule, they overestimate the cutoff latitudes for protons with energies of tens of MeV. The discrepancies increase fast with enhanced geomagnetic activity. Discussing that problem, *Smart et al.* [2000] mention that the high-latitude cutoff is very dynamic. This is due to variations in the coupling of the internal and external magnetic fields as the Earth rotates within a magnetosphere that is oriented in a solar wind flow coordinate system. Furthermore, propagation of a low-energy charged particle to the polar regions through the magnetospheric tail is very complicated due to high variability of that magnetospheric region. Another important factor is geomagnetic disturbances, which affect substantially the configuration of whole magnetosphere. As a result, magnetospheric models have difficulties in predicting the geomagnetic field, especially during magnetic storms.

The geomagnetic cutoffs are measured in space (satellite-based) experiments at low-altitude satellites with polar orbits. In practice, the cutoff thresholds $\Lambda$ are determined for one or a few rigidities of SEP. *Ogliore et al.* [2001] studied cosmic ray nuclei in the range $500 < R < 1700$ MV detected by MAST instrument onboard the SAMPEX satellite at altitudes of ~600 km. They found the following best fit for the high-energy protons (HEP) under quiet conditions:

$$R(\text{GV}) = 15.062 \cos^4(\lambda) - 0.363 \qquad (2).$$

Observed cutoff latitudes are revealed to be lower than those calculated from the dipole approach, which means that particles with very low rigidity, moving along the field lines of the magnetotail lobes, are able to penetrate in the polar region and occupy the area of polar cap (PC).

*Leske et al.* [2001] studied SAMPEX data on 8-15 MeV/nuc He nuclei ($R$~250-340 MV equivalent to ~30 - 60 MeV protons) and determined a dependence of the cutoff latitudes from the storm-time *Dst* index as $\lambda = 64.2 + \text{Dst}/19.11$. *Birch et al.* [2005] studied SEP protons with energies of 35 - 70 MeV detected by NOAA satellites POES-15 and POES-16 at ~850 km altitude and obtained a weaker dependence on *Dst*: $\lambda = 63.6 + \text{Dst}/35.7$. They also revealed a significant day-night asymmetry, such that the cutoff latitude being on average about 1.5° higher at noon than in the night period. With SAMPEX data, *Smart and Shea* [2003] showed that the cutoff latitude for 29 - 64 MeV protons decreases from quiet-time $\lambda = 65°$ at $Kp = 1$ to disturbed $\lambda = 60°$ at $Kp = 6$. We can see that for the protons in similar energy range of >30 MeV, different authors obtain different dependencies on geomagnetic activity.

Protons with lower energies reveal more complicated dynamic. *Fanselow and Stone* [1972] studied geomagnetic quiet cutoff for protons in six energy ranges from 1.28 MeV to 24.4 MeV. They found a strong day-night asymmetry of cutoff latitudes, especially for lower energies. Namely, protons with energies below 2 MeV are cut off at $\lambda \sim 71°$ to $72°$ in the dayside and at $\lambda \sim 65°$ in the night side. The asymmetry decreases with increasing energy such that for ~24 MeV protons, the cutoff latitude is found to be at ~66° in the dayside and at ~64° in the night side. *Ivanova et al.* [1985] reported even more complex behavior of the cutoff for ~1 MeV protons. The threshold latitude is characterized by prominent day-night and dawn-dusk asymmetries: at nighttime and dusk protons penetrate at lower latitudes of ~67° and ~70°, respectively, than at dayside and dawn where the threshold is located at ~74°. The cutoff latitude demonstrates a strong nonlinear dependence on the *Dst* index, auroral electroject (*AE*), and solar wind dynamic pressure.

Significant difference of the high-latitude geomagnetic field from the dipole causes very complex geometry and dynamics of geomagnetic cutoffs for energetic electrons. Because of very small rest mass (511 keV), the SEP electrons with typical energy of ~100 keV have rigidity of ~0.33 MV, which is equivalent to very low proton energy of ~0.05 keV. For that rigidity, the dipole approach [see *Smart and Shea*, 1994] gives the cutoff latitude of $\lambda \sim 86°$, which is much higher than $\lambda \sim 72°$ observed for the >30 keV electrons at low-altitude satellite Intercosmos-17 [*Gotselyuk et al.*, 1990]. *Antonova et al.* [1989] demonstrate that the cutoff boundary for >30 keV electrons is located below the high-latitude edge of auroral precipitations and, hence, the SEP electron fluxes can be substantially contaminated by the magnetospheric electrons. That makes it difficult to study the cutoff boundary for the electrons with energies below ~100 keV.

Numerous studies reveal a prominent day-night asymmetry of electron fluxes at high latitudes [e.g. *Frank et al.*, 1964; *Williams and Mead*, 1965; *McDiarmid et al.*, 1975; *Gotselyuk et al.*, 1998]. For electrons with energies of hundreds of keV, the magnitude of asymmetry is estimated of ~4°, and electron fluxes penetrate to lower latitudes at nightside. The day-night asymmetry mainly results from geomagnetic field distortion caused by the magnetotail current sheet. In addition, *Sergeev et al.* [1987] found that in the nightside, SEP electrons precipitate in the region of closed magnetic field lines. They showed that so deep penetration to the magnetosphere was due to the non-adiabatic scattering of electrons in the magnetotail current sheet. As a result, SEP electrons with very low rigidities are able to access much lower geomagnetic latitudes than those predicted from the adiabatic motion in the magnetosphere.

The electron trajectory is very sensitive to the configuration of geomagnetic field. *Antonova et al.* [1989] found not only the day-night but also a dawn-dusk asymmetry of the cutoff boundary for >30 keV electrons. For geomagnetic quiet, the cutoff boundary is skewed by ~0.5° towards dusk. This effect is attributed to the local time asymmetry in distribution of the field-aligned currents. The asymmetry of quiet magnetosphere can be related to asymmetrical conditions in the solar wind flow around the magnetosphere and/or to asymmetry of the ionospheric sources of ions in the magnetosphere.

We can see that tracing low-energy protons and energetic electrons is a very complicated problem because of complex geomagnetic field structure at high latitudes. Hence, empirical modeling becomes an important tool for studying penetration of energetic protons and electrons into the magnetosphere. In this paper we perform a detailed study of solar energetic particles observed by low-latitude NOAA POES satellites during long-lasting SEP events in December 2006 (Section 2). An empirical model of cutoff boundaries for the SEP in a wide range of rigidities and under various geomagnetic conditions is developed in Section 3. In Section 4, the model is applied for the prediction of polar cap absorption effects observed by high latitude ionosondes. Results of the study are discussed in Section 5 and Section 6 is conclusion.

## 2. Experimental data

Energetic particle measurements are acquired from five NOAA Polar Orbiting Environmental Satellites (POES): NOAA-15, NOAA-16, NOAA-17, NOAA-18, and METOP-02 (hereafter P15, P16, P17, P18 and P02, respectively). The POES satellites have polar Sun-synchronous orbits at altitude of ~800 - 850 km (~100 min period of revolution). On average, a POES satellite passes the polar region in each hemisphere every 20 min, which provides an opportunity to conduct high-resolution observations of the SEP penetration to the polar cap.

Onboard the POES spacecraft, the energetic particles are measured by a Medium Energy Proton and Electron Detector (MEPED), which constitutes the Space Environment Monitor (SEM) [see *Huston and Pfitzer*, 1998; *Evans and Greer*, 2004; also the latest version at http://poes.ngdc.noaa.gov/docs/sem2_docs/2006/SEM2v2.0.pdf].

A set of solid-state energetic particle detectors (MEPED) is designed for monitoring the energetic electrons and low-energy protons (LEP) over a range extending from 30 keV to >6.9 MeV. The telescope distinguishes particles with pitch angles of ~0° and ~90° (moving along and perpendicular the magnetic field lines). The nominal geometric factor of the telescope is 0.01 cm$^2$ sr. The protons are detected in six energy bands: 30 to 80 keV, 80 to 240 keV, 240 to 800 keV, 800 to 2500 keV, 2.5 – 6.9 MeV, and >6.9 MeV. Note that the front silicon solid-state detector in the proton telescope suffers radiation damage that becomes significant after two to three years of operation. That effect can effectively increase the LEP energy thresholds for the P15, P16 and P17 satellites launched before the year 2004.

Electrons are measured by MEPED in three energy bands: >300 keV, >100 keV, and >300 keV. Note that electron detectors are also sensitive to protons with energies of >210 keV, which are able to pass through the passive shield of the detector. The correction of proton contamination is difficult due to the radiation damage effect in the proton telescope.

The SEM instrument package includes also a set of four separate omni-directional solid-state detectors, which monitor the HEP fluxes in four integral channels: >16 MeV, >36 MeV, >70 MeV, and >140 MeV. In the case of SEP, we suppose that the high-energy protons arrive at the polar cap (PC) from the upper hemisphere, i.e. from zenith angles below ~90°. In this case the geometric factors for the high-energy channels are 0.849, 0.849, 0.182 and 0.182 (cm$^2$ sr), respectively.

Examples of particle fluxes observed by P17 in the polar cap are presented in Figure 1. Note that P17 moved from nightside to dayside. During the SEP event, the fluxes exceeded substantially the background level, detected at 2035 –2105 UT on December 4. During quiet days, the background fluxes in the PC are produced by secondary particles generated in interactions of relativistic galactic cosmic rays with the atmosphere. As one can see from the comparison of $L$-shell time profiles, the location of satellite during the quiet interval is practically the same as that during the SEP event at 2130 - 2200 UT on December 7. The $L$-shells are calculated from IGRF geomagnetic field model of epoch 2005. Similar to previous studies [*Ogliore et al.*, 2001; *Birch et al.*, 2005], we determine the cutoff latitude for the SEP as invariant latitude at which the flux is half of its mean value calculated at $L$ >15 (geomagnetic latitudes >75°). The uncertainty of this method is estimated of about a few tenths of degree [*Leske et al.*, 2001].

The omni-directional flux of HEP with $E$ > 36 MeV (Figure 1a) demonstrates symmetrical location of the cutoffs at $L$ ~ 4.2 in the day and night sides. LEP with $E$ = 2.6 - 6.9 MeV (Figure 1b) are cut off at higher latitudes ($L$~6.8) in dayside compared with $L$ ~ 4.8 in nightside. More prominent day-night asymmetry is found for the SEP electrons (Figure 1c,d), which are cut off at $L$ ~ 8.5 and ~5.5 in day and night side, respectively.

It is rather complicated to determine cutoff boundaries for electrons because of the intense precipitation in the auroral region and from the outer radiation belt (ORB). In Figure 1 the ORB is easily revealed as a region of very high fluxes of trapped and quasi-trapped electrons with pitch angles of ~90° at $L < 6$ in dayside. Less intense fluxes of electrons precipitating from the ORB (pitch angle ~0°) can also be found in that region, especially for energy range of 100 keV and below. Auroral precipitations are enriched by energetic electrons with pitch angles close to 90°. Those electrons reach mirror points at altitudes of ~850 km. It is important to note that the fluxes of SEP electrons are almost isotropic. This property permits us to distinguish the SEP electrons from the electrons of magnetospheric origin. However, during strong auroral activity the fluxes of precipitating electrons become very intense and occupy a wide range of latitudes that results in ambiguous determination of the cutoff latitudes for SEP electrons. The fluxes of electrons with energies <100 keV are substantially contaminated by intense auroral precipitations such that determination of the cutoff latitudes for them is impossible in numerous cases.

Figure 2 show mean fluxes of energetic particles observed by each of five POES satellites on 5 to 17 December 2006 in the northern PC at $L > 15$. SEP events are associated with strong solar flares. Four X-class solar flares originated from an active region AR 930 were observed by GOES-11 during that time. The first flare at ~1030 UT on December 5 occurred near eastern limb and was followed by slowly growing electron fluxes and by a small enhancement of HEP. The second limb flare at ~1900 UT on December 6 resulted in additional growth of those fluxes and rapid enhancement of the LEP and helium fluxes. Note that the helium fluxes are much (>10 times) weaker than the proton fluxes. Particles of different energies reached maximum at different time on December 7 to 8, because of velocity dispersion. The time profiles of SEP observed on 5 to 12 December are typical for the eastern solar flares. Those SEP events are accompanied by moderately disturbed geomagnetic activity with $Kp < 5$.

The flare at ~0200 UT on December 13 was almost central and characterized by pulse injection of intense fluxes of particles, especially HEP. During the following day, fluxes of LEP and energetic electrons grew dramatically and reached maximum at ~14 UT on December 14, when a CME-driven interplanetary shock (IS) impacted on the Earth's magnetosphere. The IS was effective in accelerating the electrons, LEP and He ions such that their fluxes were gradually increasing and peaked at the time of the shock arrival. The time interval preceding the shock is characterized by relatively quiet geomagnetic activity with $Kp < 3$.

The CME caused a severe geomagnetic storm with maximum $Dst$ variation of ~-150 nT and $Kp \sim 8+$ observed after midnight of December 15. The fourth X-class solar flare occurred in the beginning of the storm main phase at ~2300 UT. However, HEP fluxes from that flare were modest. The LEP, helium, and energetic electrons practically did not respond to the flare.

In Figure 2 we find a spread in the proton fluxes measured by different POES satellites. The spread is mostly noticeable during slow variations in the fluxes. In Figure 3 we compare POES measurements of the 2.5-6.9 MeV and >36 MeV protons near their maximum fluxes at ~0000 UT on December 8. We can see that all satellites, except P16, detected almost same proton fluxes. For P16 we find a ~20% underestimation of the LEP intensity and ~5% overestimation of the HEP intensity. Latter effect may be due to different calibration of the omni-directional detector on P16. Actually, that effect is small and we can neglect it.

The underestimation of LEP fluxes by P16 is probably due to the radiation damage effect, which raises the LEP energy thresholds. The influence of that effect to the accuracy of cutoff latitude determination depends on the LEP spectrum. Figure 4 shows an average integral spectrum of solar protons measured in the polar cap by the POES satellites at 0000 – 0100 UT on December 8, 2006. The spectrum of LEP (with energies below ~16 MeV) can be fitted by a logarithm function of energy: $F(>E) = a + b\log(E)$. Using the logarithmic approach, we can estimate a shift of energy threshold $\Delta E$, which leads to ~20% decrease of the $2.5 - 6.9$ MeV proton flux measured by P16. The flux in a given energy range from $E_1$ to $E_2$ of protons with logarithmic integral spectrum can be presented as:

$$dF = F(> E_1) - F(> E_2) = b(\log(E_1) - \log(E_2)) \qquad (3)$$

We assume that the lower and upper thresholds increase by the same amount of energy $\Delta E$. Hence, the equation (3) can be rewritten as: $0.8 \cdot dF = b(\log(E_1 + \Delta E) - \log(E_2 + \Delta E))$. Taking the measured flux of 4500 $(\text{cm}^2 \text{ s sr})^{-1}$ and the best fit of $b = 4433$, we estimate the $\Delta E$ of ~1 MeV. From equation 2 we can calculate that the rise of energy threshold from 2.5 to 3.5 MeV leads to only a little decrease of the cutoff latitude from ~65.7° to ~65.5°. Similar change of a few tenth of degree was found for other energy thresholds of LEP. This change is within the experimental error of cutoff latitude determination. Hence, this effect can also be neglected.

Other errors can be attributed to contamination of the LEP by fluxes of heavier ions, HEP and relativistic electrons (>2 MeV). Those particles are able to penetrate to lower latitudes that result in underestimation of the cutoff latitudes for the LEP. In Figure 2 we have shown that the helium fluxes, detected by SOHO in the interplanetary medium, are negligibly small. Ions with higher mass have even weaker fluxes. During strong solar flares, accompanied by fast interplanetary shock, electrons can be effectively accelerated up to relativistic energies [*Zhang et al.*, 2003; *Mewaldt et al.*, 2005]. Relativistic electrons with $E$ >3 MeV contribute to the count rate of the LEP channels. On December 2006, solar flares were not very strong and the shock was not very fast (propagating time was >24 hours). The fluxes of relativistic electrons ($E = 2.8 - 4$ MeV) detected by high-energy telescope (HET) onboard STEREO satellites (http://www.srl.caltech.edu/STEREO/Level1/HET_public.html) were >10 times weaker than the LEP fluxes. Note that until December 15 the STEREO satellites were moving between the Earth and Moon orbit.

It is rather difficult to avoid the radiation effects caused by the high-energy and relativistic protons. In order to eliminate those effects, we select only time intervals when the fluxes of LEP and electrons are stronger than those of HEP. As a result, for LEP we choose time intervals from 1100

UT on December 7 to 1900 UT on December 10 and from 1500 UT on December 13 to 2200 UT on December 14. The electrons are analyzed from 1200 UT on December 7 to 2100 UT on December 9 and from 1900 UT on December 13 to 2200 UT on December 14.

## 3. Cutoff boundary modeling

Figure 5 shows scatter plots of determined cutoff latitudes in a coordinate system of invariant co-latitude $\Lambda$ versus MLT. The invariant co-latitude is defined as $\Lambda = 90 - \lambda$, where $\lambda$ is invariant latitude related to the $L$-shell as $L = \cos^{-2}(\lambda)$. This coordinate system has the origin in the magnetic pole. One can see that the experimental points cover practically all sectors of MLT. Though, some MLT intervals void of data because of Sun-synchronous orbits of the POES satellites. The cutoff latitudes spread widely (>10°) towards higher $\Lambda$, i.e. lower latitudes. This spread is originated from geomagnetic activity varying significantly in December 2006 (see Figure 2).

The numbers of cutoff latitudes determined for each of the various SEP species measured are listed in Table 1. Because we use 5 satellites, the resultant statistics involves hundreds of points that allow multi-parametric analysis and modeling of the cutoff latitudes versus MLT, i.e. cutoff boundaries. In the pure dipole approach, the cutoff boundary should be a circle with a certain radius equal to the cutoff latitude $\Lambda$, which does not depend on MLT. That approach is pretty good for the HEP, whose penetration is controlled mainly by the inner magnetosphere at $L < 6$. However, this is not the case for LEP and energetic electrons because penetration of those particles depends substantially on the configuration of outer magnetosphere at $L > 10$.

In general, the shape of cutoff boundary can be described by an ellipse, which is tilted relative to the dawn-dusk line and shifted from the origin, as shown in Figure 6. For modeling we use a coordinate system of $\Lambda$ versus rotated magnetic local time (rMLT) calculated in degrees from the X-axis pointed dawnward. The Y-axis is pointed to the noon. The ellipse is defined by 5 parameters: major *as* and minor *bs* semi-axes, location of the center (*X*o, *Y*o) and azimuth angle *φ* of the major semi-axis. Hence, fitting a cutoff boundary by ellipse requires at least 5 experimental points, i.e. at least three passes of the POES satellites above the polar cap. On average, those passes occur every hour. We use 2-hour interval with running step of 1 hour. Within 2 hours we can collect up to 13 points for fitting the cutoff boundary for a given sort of SEP. This permits us to eliminate data gaps and smooth sharp variations. The 2-hour interval (almost 6 passes) permits also to average out the non-uniform coverage of the polar cap by the POES satellites, i.e. some satellites only scratch the polar cap while others pass right through the polar cap.

### 3.1. Geomagnetic quiet cutoffs

In Figure 7 we present the cutoff boundaries determined for quiet geomagnetic conditions (*Kp* ~ 0 and *Dst* ~ 0) at 1100 – 1300 UT on December 9. Those boundaries can be well fitted by ellipses. It is interesting to note that the cutoff boundaries for the LEP and electrons in the northern hemisphere are slightly different from those in the southern hemisphere. Perhaps, this difference is related to the different configuration of high-latitude geomagnetic field. During solstice, the geo-dipole is mostly tilted in the noon-midnight meridional plane that results in a significant north-south asymmetry of the outer magnetosphere. In December, the southern and northern PCs are tilted sunward and anti-sunward, respectively. As a result, the southern part of magnetosphere is more affected and deformed by the solar wind than the northern one. Because of this deformation, the cutoff boundaries in the southern hemisphere are more elongated along the noon-midnight line.

The cutoff latitudes for various rigidities vary with local time in different and pretty complicated manners. Using the elliptical approach, we can study this complex behavior as a dependence of the ellipse parameters on the rigidity. The rigidities are calculated for the lower thresholds of MEPED energy bands (see Table 1).

Figure 8 shows a gradual increase of the semi-axes with the rigidity $R$. Here we take into account of the fact that invariant latitude $\lambda$ is related with invariant co-latitude $\Lambda$ as: $\cos^4(\lambda) = \sin^4(90-\lambda) \equiv \sin^4(\Lambda)$. For convenience, we will indicate the values of both major and minor semi-axes as $\Lambda_S$. Note that $\Lambda_S$ is calculated from the center of ellipse, which is not necessarily coincide with the magnetic pole, the origin of coordinates. In HEP range with rigidities above 100 MV ($E > 6$ MeV), quiet-time major semi-axes (hereafter $as_q$) in the southern hemisphere approach the average fitting obtained by *Ogliore et al.* [2001] (see also Equation 2) for slightly disturbed geomagnetic conditions with $Kp \sim$ 2+. In our case of geomagnetic quiet ($Kp \sim 0$), the cutoff latitudes $\lambda$ are expected to be higher, i.e. $as_q$ should be smaller. Indeed, in Figure 8 the major semi-axes in the northern hemisphere and the quiet-time minor semi-axes (hereafter $bs_q$) in both hemispheres are slightly smaller than those predicted by *Ogliore et al.* [2001]. In the HEP range, the dependence for $\sin^4(\Lambda_s)$ can be presented as a linear function of rigidity $R$:

$$\sin^4(\Lambda_S) = A + B \cdot R \quad (4)$$

The best-fit coefficients $A$ and $B$ obtained for the major and minor semi-axes are listed in Table 2.

The cutoff latitudes for LEP and electrons with rigidities below 100 MV demonstrate different behavior: the semi-axes $as_q$ and $bs_q$ decrease exponentially with decreasing rigidity. It seems that penetration of those particles to the inner magnetosphere is impeded such that the low-energy protons and electrons were unable to penetrate to lower latitudes in the dayside while the magnetotail in the nightside is accessible in a wide range of latitudes, as we can see in Figure 7. However, the cutoff latitudes for LEP and electrons are still higher than those derived by *Smart and Shea* [1994] from the pure dipole approach. In the whole range of proton rigidities, dependence of $\Lambda_S$ on rigidity $R$ can be presented by the following expression:

$$\sin^4(\Lambda_S) = A + B \cdot R + C \exp(DR) \quad (5)$$

Here coefficients $C$ and $D$ describe the exponential increase of $\sin^4(\Lambda_s)$ with rigidity in LEP range. At high rigidities, equation (5) approaches asymptotically to equation (4). The best-fit values of coefficients $C$ and $D$ are listed in Table 2.

A scatter plot of the coordinates $X$o and $Y$o of cutoff boundary centers versus rigidity is shown in Figure 9. As one can see, the coordinate $X$o is almost 0°, i.e. the centers are located very close to the noon-midnight meridian. The coordinate $Y$o is negative and its value increases with decreasing rigidity such that the cutoff boundaries for the LEP and electrons are shifted tailward by several degrees. This shift is mostly prominent at lowest rigidities R<100 MV. In HEP range, the shift still persists but its value is very small (<1°). From Figures 7 and 9, we can conclude that during geomagnetic quiet period, LEP and energetic electrons are characterized by a prominent day-night asymmetry such that the centers of cutoff boundaries are substantially shifted tailward.

Figure 10 demonstrates a scatter plot of the azimuth angle $\varphi$ (between the major semi-axis and $X$-axis) versus rigidity. Within the errors, the angle is practically equal to zero for protons. For the electrons, especially in the southern hemisphere (see Figure 7), the cutoff boundary is elongated along the noon-midnight meridian such that the azimuth angle approaches to ~-70°. This is an additional demonstration of the difference in penetration of the LEP and electrons to the outer magnetosphere.

*3.2. Cutoff variations with geomagnetic activity*

Cutoff boundaries for SEP are significantly modified during geomagnetic disturbances such as substorm-related auroral activity and geomagnetic storms. Figure 11 shows the SEP cutoff boundaries determined during storm main phase at 1700 UT on December 14. Comparing with the geomagnetic quiet (Figure 7), one can see a significant difference in the shape and size of cutoff boundaries. As expected, SEP penetration region extends to lower latitudes (larger $\Lambda$). We can also find prominent tailward and duskward shifting of cutoff boundaries. Fitting of storm-time cutoff boundaries has a large error because of wide spreading of cutoff latitudes determined within 2-hour interval. This spreading is owing to the fast changes in the geomagnetic field due to the storm-time magnetospheric current systems with characteristic time of tens of minutes.

Time profiles of the geomagnetic indices and fitting parameters of the cutoff boundaries are presented in Figure 12. The elliptical parameters are characterized by very high variability in various time scales. Sharp changes within one-two hours can be attributed to a noise due to fast variations of statistics, which are caused by quiet often data gaps and different coverage of the polar cap by different POES satellites. Those "uncertainties" in determination of the cutoff boundaries affect substantially the coordinates of center $Xo$ and $Yo$ because they are very sensitive to the number and distribution of fitting points along the boundary. In case of good statistics (see Figure 11) we observe smooth changing of the elliptical parameters. Poor statistics (see left panel in Figure 7) is accompanied by enhanced noise.

In Figure 12 one can clearly see long-term variations with duration from several hours to days. It is important to note that the amplitudes of long-term variations exceed the amplitude of noise. That makes possible to model the long-term variations by means of multi-parametric regression. The semi-axes ($as$ and $bs$) and tailward shift (coordinate $Yo$) demonstrate co-variation with the geomagnetic indices. On average, their values are smaller during weaker geomagnetic activity (smaller $Kp$). We can also find a detailed correspondence between short-time enhancements of the parameters and peaks in the $AE$ and $Dst$ indices. The amplitude of variations depends on the rigidity: LEP and electrons are more sensitive to the geomagnetic activity.

The dawn-dusk shift ($Xo$) varies about zero with amplitude of ~1°. In HEP range, the value of $Xo$ is predominantly positive (negative) in the northern (southern) hemisphere. It seems that the amplitude of $Xo$ for HEP anti-correlates with the amplitude of geo-dipole tilt angle PS. The amplitude of $Xo$ increases with geomagnetic activity. In the case of LEP and electrons, dynamics of $Xo$ is different. On average, $Xo$ is close to zero or positive for weak geomagnetic activity. During enhanced activity, $Xo$ tends to be more negative (duskward shifting) in both hemispheres.

Most prominent variations are revealed during geomagnetic storm on December 14 to 15, 2006. All the parameters demonstrate a substantial and fast growth during the storm main phase lasting from ~1200 UT to ~2400 UT on December 14. The amplitude of growth reaches up to ~10° for the semi-axes and of ~3° and ~5° for $X$o and $Y$o, respectively. Then on December 15 the parameters are recovering but in a different manner. The semi-axes are decreasing slowly and following to the gradual positive $Dst$ variation at storm recovery phase. The coordinates $X$o and $Y$o are recovered fast during the storm maximum at ~0000 UT on December 15.

The complex dynamics of cutoff boundaries is modeled by using a multi-parametric linear regression. In the previous section we found that the dependence of semi-axes on rigidity for protons is strongly nonlinear. In order to eliminate that nonlinearity, we model the residual semi-axes $\Delta as$ and $\Delta bs$ for the protons:

$$\Delta as = as - as_q, \qquad (6a)$$

$$\Delta bs = bs - bs_q. \qquad (6b)$$

We model an elliptical parameter $P$ (i.e. $\Delta as$, $\Delta bs$, $X$o, $Y$o, or $\varphi$) separately in the northern and southern hemisphere as a linear superposition of geomagnetic indices ($Dst$, $Kp$, $AE$) and geo-dipole tilt angle PS:

$$P = P_0 + P_{Dst} \cdot Dst + P_{Kp} \cdot Kp + P_{AE} \cdot AE + P_{PS} \cdot PS \qquad (7)$$

Here $P_0$, $P_{Dst}$, $P_{Kp}$, $P_{AE}$, $P_{PS}$ are regression coefficients, $Dst$ and $AE$ indices are expressed in nT, tilt angle $PS$ in degrees. For the $Kp$ index we use a decimal representation, i.e $Kp*10$. For instance, $Kp = 1+$ is represented as 13.33, $Kp = 6-$ is 56.67 etc.

At the first step, we use a linear regression in order to calculate the coefficients $P_0$, $P_{Dst}$, $P_{Kp}$, $P_{AE}$, and $P_{PS}$ in each energy band of protons. For each elliptical parameter we determine a meaningful set of regression parameters by exception one or other index from equation (7) and checking the corresponding change of regression error. The error increases abruptly when a meaningful parameter is excluded from the regression. By this way we find the following sets of parameters providing a reasonable accuracy of the regression: ($Dst$, $Kp$) for the $\Delta as$, $\Delta bs$, and $Xo$; ($Dst$, $Kp$, $PS$) for the $Yo$. The azimuth angle $\varphi$ does not reveal any meaningful dependency.

At the second step, we use the regression coefficients, obtained in the different energy bands, in order to study their dependence on the proton rigidity. Regression coefficients $P_{Dst}$ and $P_{Kp}$, calculated for the major and minor semi-axes at various proton rigidities, are presented in Figure 13. One can see that the coefficients can be well fitted by a logarithmic function of rigidity:

$$P_i = A_i + B_i \log R \qquad (8)$$

The corresponding coefficients $A_i$ and $B_i$ are presented in Table 3. Here we use a natural logarithm such that $\exp(\log R) = R$. The negative values for $P_{Dst}$ and positive for $P_{Kp}$ imply a growth of the semi-axes with geomagnetic activity. This growth is larger for the lower rigidities (energies), where the absolute value of regression coefficients is higher.

In Figure 13 one can find a lack of regression coefficients $P_{Dst}$ and $P_{Kp}$ for the protons with rigidity 70 MV and 117 MV, respectively. Those coefficients were found very much different from the vast majority of points. The large difference is probably resulted from a relatively small statistics collected for those energy bands (see Table 1). In addition, cutoff boundaries for those protons are characterized by very complicated dynamics as one can see in Figure 12. It is unlikely that the penetration of protons with rigidities of 70 MV and 117 MV is very much different from that of one found for the protons with neighbor rigidities. Hence, we suppose that those protons follow the general rules and, thus, we can determine their regression coefficients $P_{Dst}$ and $P_{Kp}$ by using a linear interpolation.

The coordinates $Xo$ and $Yo$ of center of cutoff boundaries have more complex dynamics. Dependence of their regression coefficients $P_0$, $P_{Dst}$ and $P_{Kp}$ on rigidity is presented in Figure 14. As we have demonstrated for geomagnetic quiet period, the coordinate $Xo$ is small and does not depend on the rigidity, while the $Y$o is negative and grows with decreasing rigidity. This dependence can be fitted by a linear function of the logarithm of rigidity (equation 8), as one can see in Figure 14(a).

The absolute value of regression coefficients $P_{Dst}$ and $P_{Kp}$ for the coordinates $X$o and $Y$o approaches to 0 with increasing rigidity. That is in good agreement with the fact that the penetration of high-energy SEP is less sensitive to the geomagnetic disturbances. Enhanced geomagnetic activity is characterized by large positive $Kp$ and negative $Dst$ variation. Hence, negative values of the $P_{Kp}$ and positive $P_{Dst}$ for the $X$o imply a growth of the duskward ($X$o < 0) shifting with increasing geomagnetic activity.

Variations of the $Y$o with geomagnetic activity are not so obvious: both $P_{Dst}$ and $P_{Kp}$ are negative. The negative values of coefficient $P_{Kp}$ imply a negative (tailward) shift of the $Y$o with growing magnetic activity. In contrast, during magnetic storm with large negative $Dst$ variation, we find an opposite effect associated with negative values of the coefficient $P_{Dst}$ that result in a shift of the coordinate $Y$o toward noon.

The location of cutoff boundaries for protons reveals a dependence of the geo-dipole tilt angle $PS$. This dependence for the coordinate $X$o is relatively weak and ambiguous, as we have found in Figure 12. Instead, the linear regression reveals a substantial $Y$o variation with the tilt angle, as shown in Figure 15. In the LEP range, the amplitude of variation is estimated to be ~1° that is beyond of the methodic error (???). The amplitude decreases exponentially with increasing rigidity, such that the variation practically diminishes at rigidities above 200 MV.

Negative values of both the regression coefficient $P_{PS}$ and the tilt angle $PS$ result in a positive (i.e. sunward) shift of the cutoff boundaries in both hemispheres. The sunward shift increases with the amplitude of tilt angle. Such behavior can not be explained by simple geometrical effects, which should be opposite in different hemispheres. It might be a manifestation of the equinoctial effect, which consists in higher ability of the magnetosphere to be disturbed for smaller tilt angles [*O'Brien* *and McPherron*, 2002]. Hence, at higher tilt angles the average tailward (sunward) shift appears smaller (larger).

Variations of the azimuth angle $\varphi$ with geomagnetic activity for various rigidities are very complex and irregular. We could not find any meaningful dependency from the geomagnetic indices. Hence, the azimuth angles were averaged for each range of rigidities. In Figure 16 we can see that the mean $\varphi$ increases linearly with the logarithm of rigidity: from $\sim$-20° in LEP range to $\sim$0° in HEP range. Perhaps, this result indicates a tendency of the cutoff boundaries to have negative azimuth angles during geomagnetic disturbances, as demonstrated in Figure 11.

We find that the regression coefficients can be expressed as linear functions of the logarithm of proton rigidity (equation 8) and, hence, the general expression for the parameters $\Delta as$, $\Delta bs$, $X$o, $Y$o, and $\varphi$ can be written as the following:

$$P = A_0 + B_0 \log R + (A_{Dst} + B_{Dst} \log R) \cdot Dst + (A_{Kp} + B_{Kp} \log R) \cdot Kp + (A_{PS} + B_{PS} \log R) \cdot PS \quad (9)$$

Note that we do not find meaningful dependency for the *AE* index. The best-fit values of coefficients $A_i$ and $B_i$ are presented in Table 3. By this way we have constructed a dynamical model of the cutoff boundaries for the protons in energy range from 240 keV to >140 MeV. The cutoff invariant co-latitude $\Lambda$ at given local time rMLT is calculated as a point at elliptical boundary, which is determined by the rigidity $R$ and geomagnetic parameters *Dst*, *Kp*, and *PS*. The expressions for the calculation of $\Lambda$ are following:

$$\Lambda = \sqrt{x^2 + y^2} \qquad (10a)$$

$$x = Xo + as \cos rMLT \cos \varphi - bs \sin rMLT \sin \varphi \qquad (10b)$$

$$y = Yo + as \cos rMLT \sin \varphi + bs \sin rMLT \cos \varphi \qquad (10c)$$

The values of *as*, *bs*, *X*o, *Y*o, and $\varphi$ are determined from equations (5), (6), (9) with coefficients presented in Tables 2 and 3.

Comparison of the model predictions with the cutoff invariant co-latitudes determined for the protons is presented in Figure 17a. The common statistics, collected in all energy ranges, consist of 4768 and 4728 points in the northern and southern hemisphere, respectively. The model accuracy is represented by a root mean square deviation (RMSD) and mean relative error (MRE):

$$\text{RMSD} = \sqrt{\sum \left( \Lambda_{\text{mod}}^2 - \Lambda_{obs}^2 \right) / N} \qquad (11a)$$

$$\text{MRE} = \frac{100\%}{N} \sum \frac{\left| \Lambda_{\text{mod}} - \Lambda_{obs} \right|}{\Lambda_{obs}} \qquad (11b)$$

where $\Lambda_{\text{mod}}$ and $\Lambda_{obs}$ are, respectively, modeled and observed (i.e. determined) latitudes and $N$ is number of observations. For the model we obtain the RMSD (MRE) of 1.3° (4.2%) and 1.4° (4.1%), respectively, in the northern and southern hemisphere.

In case of SEP electrons, we model only cutoff boundaries for the >300 keV energy range. The fluxes of electrons with lower energies are significantly contaminated by intense auroral precipitations such that determination of the cutoff latitudes is difficult in most cases. Very intense auroral activity is accompanied sometimes with precipitation of >300 keV electrons at high latitudes. Those cases have been eliminated.

The elliptical parameters $as$, $bs$, $X$o, $Y$o, and $\varphi$ of the cutoff boundaries are considered as a linear function of geomagnetic indices (see equation (7)) separately in northern and southern hemisphere. For each parameter, we calculate regression coefficients. Similar to protons, we select the meaningful set of regression parameters. As a result, we have developed a model of cutoff invariant co-latitudes for >300 keV electrons. The model is described by equations (7) and (10) with coefficients presented in Table 4.

Comparing Table 4 with Table 3, we find that coefficients for some regression parameters are similar. For instance, the major ($As$) and minor ($bs$) semi-axes as well as duskward shifting ($X$o) have similar behavior for the electrons and protons: they increase with geomagnetic activity. The coordinate $Y$o for electrons and protons has similar dependence on the tilt angle $PS$. However, we can also reveal some differences. The coordinate of center *Yo* is independent of *Kp* and *Dst*, but proportional to *AE*. Such dependence might indicate importance of intensified field-aligned currents, which magnetic effect probably prevents the penetration of SEP electrons to low latitudes at nighttime.

The azimuth angle $\varphi$ demonstrate a dependence on the geomagnetic parameters *Dst* and *PS*. Note that the coefficients $P_0$, $P_{Dst}$ and $P_{PS}$ have different signs in different hemispheres. Hence, that might be a geometric effect related to opposite dynamics of the geo-dipole axis. Indeed, a variation of the tilt angle from –10° to –30° is accompanied by the opposite changes of azimuth angle in northern and southern hemisphere, respectively, from ~0° to -30° and from -20° to ~-7°. The same anti-correlation we find for the dependence on the *Dst* index.

Figures 17c and 17d demonstrate comparison of the model predictions with the observations for electrons. The model RMSD (MRE) are obtained of 2.8° (12%) and 2.5° (9.4%), respectively, in the northern and southern hemisphere. We should note that the model accuracy is moderate because of modest statistics of ~400 measurements (see Table 1). Hence, the dynamics of cutoff boundary for the electrons, especially the day-night asymmetry and the boundary rotation, requires further investigations based on extended statistics.

## 4. Application for polar cap absorption effect

The long-lasting SEP events in December 2006 were accompanied by effects of radio signal distortions at high latitudes. Those effects are studied using a network of Canadian Advanced Digital Ionosondes (CADIs) in high-latitude: Eureka, Resolute Bay, and Rankin [*Jayachandran et al.*, 2009]. Location of these stations are indicated in Table 5. These ionosondes record sweep frequency (1-20 MHz) ionograms every minute and make fixed frequency drift measurements every 30 seconds. Ionosondes work on the principle of radio "reflection". Reflection occurs when the transmitting frequency is equal to the local plasma frequency (condition of zero refractive index). If there are significant ionization in the D and lower E region of the ionosphere due to particle precipitation, the whole ionosonde HF signal (1-20 MHz) will be absorbed. The effect is called Polar Cap Absorption (PCA) event. PCA is a common feature of the SEP events [e.g. *Rodger et al.*, 2006].

In this study we use ionograms in order to determine the presence or absence of the reflected HF signals at given station. Figure 18 shows an example of daily variation group range at a fixed frequency for stations Eureka and Rankin on December 12, 2006. One can see that between 0100 and 0600 UT and between 0700 and 1400 UT, the ionosonde at polar cap station Eureka detects no reflected radio signals in the frequency range 1-20 MHz. The lack of reflected signal at Eureka during that period is most probably due to the absorption of radio signal by the enhanced D region ionization caused by the high-energy particle precipitation (SEP). In other words, we observe PCA event. At the same time, the ionosonde at the station Rankin, located at lower latitude, does not detect absence of reflected radio signals in the frequency range of 1-20 MHz. Hence between 0100 and 0600 UT and between 0700 and 1400 UT on December 12, the longitudinal extension of PCA is restricted somewhere between $\Lambda=17.64°$ of Rankin and $\Lambda=5.35°$ of Eureka.

Using daily ionograms from above mentioned stations, we have collected statistics of PCA effect observed by three CADI stations for each hour from 6 to 16 December 2006. The 1-hour intervals of PCA observations at different stations are presented in Figure 19. Dashed horizontal lines in the figure represent the absence of reflected signals (i.e. presence of PCA) in the frequency range of 1-20 MHz. One can see that the PCA intervals observed at high-latitude stations Eureka and Resolute Bay are pretty similar. At the lower-latitude station Rankin, the number of PCA intervals is smaller. That difference can be explained by weaker SEP ionization at lower latitudes because the cutoff boundaries for SEP are located at higher latitudes (lower $\Lambda$), between the Rankin and Resolute Bay.

We use the model of SEP cutoff boundaries in order to determine which particles are mostly responsible for the observed PCA effects. We consider both variations of the SEP intensity observed by NOAA-POES (see Figure 2) and variations of the cutoff latitudes predicted by the model at MLT of each of three CADI stations. As an example, intensities and predicted cutoff latitudes for the MLT of Ranking are shown in Figure 19 for HEP, LEP, and >300 keV electrons.

In Figure 19 we also show cutoff latitudes determined from the elliptical approach of cutoff boundaries. The elliptical approach is based on fitting parameters ($as$, $bs$, $Xo$, $Yo$, and $\varphi$) determined hourly for the SEP of various energies (see Figure 12). Using the fitting parameters, we can calculate the cutoff latitude at given MLT for each hour of SEP observations (if available). The cutoff invariant co-latitudes are calculated at MLT of Rankin for the >300 keV electrons (triangles), and protons with energies >240 keV (diamonds), >2.5 MeV (circles) and >36 MeV (crosses). The co-latitudes calculated from the elliptical approach are close on average to those calculated from the model, especially for the HEP. Most difference (of several degrees) is revealed for the electrons and >240 keV protons. The large model discrepancy for the electrons is mainly due to poor accuracy in prediction of the azimuth angle $\varphi$ that results in substantial time difference between the predicted dayside minima of the cutoff co-latitude and the observed ones.

The cutoff latitudes for the LEP and electrons vary widely with MLT such that at Rankin local noon (~19 UT) the particles are cut off below the station co-latitude of $\Lambda$=17.6. The cutoffs for protons with higher energies are practically always located at larger co-latitudes. In modeling the PCA effect, we suppose that the PCA at a given station is produced by particles from a given energy range when the invariant co-latitude $\Lambda$ of cutoff for those particles is larger than the co-latitude of station and the particle flux exceeds a certain threshold. Varying the energy range and flux threshold, we obtain various model predictions for the PCA observed at Rankin, Resolute Bay, and Eureka.

In order to evaluate the predictions, we use such statistical quantities as probability of correct predictions (*PCP*), and overestimation/underestimation ratio (*OUR*) [e.g. *Dmitriev et al.*, 2003]. Those quantities are derived from four numbers *A*, *B*, *C*, and *D*, which calculation logic is presented in Table 6. Namely, the number *A* is a number of cases when the PCA effect, observed at a given station, is predicted correctly by the model, i.e. the cutoff latitude $\Lambda_m$, modeled for the particles with a given energy and intensity, is equal to or larger than the latitude $\Lambda_O$ of station. The number $B$ ($C$) is a number of wrong predictions caused by model underestimation (overestimation) of the cutoff co-latitude. The quantity $D$ is a number of correct rejections, when the model correctly predicts no PCA effects. Apparently, the sum of $A$, $B$, $C$, and $D$ is equal to the total number $N$ of cases. The statistical quantities $PCP$ and $OUR$ are defined as the following:

$$PCP = (A + D)/N \qquad (12a)$$

$$OUR = C/(C + B) \qquad (12b)$$

The best model prediction should have highest $PCP$, and $OUR$ approaching to 0.5.

We have found that observed PCA effects are hardly predicted by the fluxes of protons with energies above 6.9 MeV. As one can see in Figure 19, the HEP fluxes enhance (decrease) much earlier than the observed starts (ends) of PCA. Among the protons, the best model prediction has been found for the energy range 2.5 – 6.9 MeV with the threshold intensity of ~100 (cm$^2$ s sr)$^{-1}$. The corresponding statistical quantities are indicated in Table 7. The >300 keV electrons do not provide accurate prediction of the PCA because of numerous cases of too low cutoff co-latitudes observed and modeled for the electrons during the PCA at Rankin.

We also study dynamics of cut-off latitude for the >100 keV electrons. In order to avoid the auroral precipitations, we analyze only time intervals with low substorm activity. Figure 20 shows cutoff latitudes and elliptical fitting of the cutoff boundary for >100 keV. The average boundary is described by the following elliptical parameters: $as = 19.8°$, $bs = 18.3°$, $Xo = 1.06°$, $Yo = -5.9°$, and $\varphi = -74.8°$. Note, the night MLT sector is poorly covered by the data and, hence, the cutoff boundary in that sector might be unreliable. However, in dayside, dawn, and dusk sectors the statistics provide reliable fitting for the average cutoff boundary.

Results of application of the >100 keV electrons for prediction of the PCA effects are presented in Figure 21. The cutoff latitudes for the protons are calculated from the model, as we have demonstrated above. For electrons, cutoff latitudes are calculated from elliptical parameters obtained either from hourly approaches, when data are available, or from the average cutoff boundary described in the previous paragraph.

We consider a total effect produced by the >2.5 MeV protons and >100 keV electrons and compare it with the PCA observed at different stations. It is supposed that PCA at a given station is produced by the sum of proton and electron fluxes when two requirements are satisfied. The first one is that the minimal invariant co-latitude $\Lambda$ of cutoffs modeled for electrons and protons should be larger than the co-latitude of station. The second requirement is that fluxes of both electrons and protons should exceed certain thresholds for corresponding particles.

Varying the threshold fluxes, we settle the best model ability for the PCA prediction with highest *PCP*, and *OUR* mostly close to 0.5. The threshold fluxes and corresponding statistical quantities are presented in Table 7. One can see that the model, based on the total effect of >2.5 MeV protons and >100 keV electrons with threshold fluxes of 100 $(cm^2 \ s \ sr)^{-1}$ and 2900 $(cm^2 \ s \ sr)^{-1}$, respectively, demonstrates better capabilities for prediction of PCA at all the stations. In Figure 21 one can clearly see that with using electrons, we decrease substantially the overestimations of cutoff co-latitudes at Rankin as well as increase the number of correct predictions at Resolute Bay. Therefore, the PCA effects, observed at high-latitudes stations, are better described by the integral effect of SEP fluxes contributed by the >2.5 MeV protons and >100 keV electrons.

## 5. Discussion

Using elliptical approach of cutoff boundaries for the SEP electrons and protons of various energies, we have developed an empirical model of the cutoff latitudes in the northern and southern hemisphere. The cutoff latitude is represented as a function of rigidity, MLT, geomagnetic parameters *Dst*, *Kp*, *AE* and dipole tilt angle *PS*. For geomagnetic quiet, the model prediction is close to that obtained by *Ogliore et al.* [2001] in the range of HEP with *E* >16 MeV. However, cutoff latitudes for the low-energy protons ($E$ < 16 MeV) and electrons are found to be much higher (i.e. at smaller $\Lambda$) than the extrapolation from HEP region as one can clearly see in Figure 8.

We have to point out that the dependence of cut-off latitude on proton rigidity, expressed by equation (5), does not fit the cutoff latitudes for electrons. The observed semi-axes of the cutoff boundaries for electrons with rigidity of ~0.6 MV ($E \sim 300$ keV) are smaller by ~2° than those predicted from equation (5). It is unlikely that so large a difference is due to experimental errors and/or uncertainty of the fitting method. This rather indicates a difference in propagation conditions for electrons and protons in the outer magnetosphere and magnetotail, which are characterized by a very wide spectrum of magnetic inhomogeneities [*Sergeev et al.*, 1987]. *Droge* [2000] showed that the mean free path of SEP has an explicit velocity dependence that leads to larger values for electrons than for protons at the same rigidity below a certain threshold of several MV. Hence the LEP, having smaller mean free path, might be scattered more effectively and, thus, occupy larger spatial region in the outer magnetosphere than the electrons with the same rigidity.

We have found a very complex dependence of the SEP penetration on geomagnetic activity and dipole tilt angle. This dependence is controlled by the configuration of magnetospheric current systems including cross-tail current, ring current, and field-aligned currents. In general, enhanced geomagnetic activity causes expansion of cutoff boundaries towards lower latitudes. The magnetospheric currents are also responsible for the day-night asymmetry and duskward shifting, which we have found to be prominent for the LEP and electrons. Our results support and supplement the previous findings [*Fanselow and Stone*, 1972; *McDiarmid et al.*, 1975; *Ivanova et al.*, 1981; *Antonova et al.*, 1989].

Modeling the centers of cutoff boundaries reveal a significant tailward shifting which increases with geomagnetic activity. This shifting can be attributed to a negative magnetic effect of intensified cross-tail current, which depress the dipole magnetic field in the nightside magnetosphere. In contrast, during magnetic storm we find a shift of the centers of cutoff boundaries towards noon.

This shift might be related to a magnetic effect of the developed ring current, which substantially reduces the geomagnetic field both at dayside and nightside. As a result, the tailward shifting related to the cross-tail current is suppressed by the strong magnetic effect of symmetrical ring current.

A prominent duskward shifting of the cutoff boundaries is observed at the main phase of magnetic storm. The duskward shifting can be related to a magnetic effect of the asymmetrical part of ring current peaked in the dusk sector at the storm main phase [e.g. *Cummings*, 1966; *Liemohn et al.*, 2001; *Dmitriev et al.*, 2004]. This magnetic effect results in a depletion of the geomagnetic field in dusk region such that the SEP are able to penetrate to lower latitudes. A superposition of the magnetic effects produced by the asymmetrical ring current at dusk and by the cross-tail current at night results in large negative azimuth angles, which are revealed for the cutoff boundaries during the magnetic storm (see Figure 11).

The model application for prediction of PCA effects, observed by high-latitude ionosondes, allowed us to reveal that absorption of radio signals are associated mainly with intense fluxes of >2.5 MeV protons. This result supports the previous findings obtained by *Clilverd et al.* [2007]. They report that the PCA effect at nighttime is rather associated with >5 MeV protons than with protons of higher energies. Moreover, we reveal an important role of energetic electrons. The best model prediction of the observed PCA effects (see Figure 21 and Table 7) is obtained for the summarized effect of >2.5 MeV protons and >100 keV electrons.

In Figure 22 we show minimal accessible heights calculated for SEP electrons and protons of various energies. The calculation was based on a model of ionization losses of charged particles in the standard atmosphere (MSIS model) at high latitudes [*Dmitriev et al.*, 2008]. One can see that the >100 keV electrons and >2.5 MeV protons stop and lose most of energy in ionization of atmosphere at heights of 75 ~ 85 km and, thus, produce ionization of the ionospheric *D*-layer.

The abundant ionization of that region can be studied by using height profiles of electron content (EC) derived from FORMOSAT-3/COSMIC radio occultation measurements [e.g. *Hajj et al.*, 2000].

During SEP events on December 2006 we observe a prominent enhancement of the EC at height below 100 km [*Dmitriev et al.*, 2008]. Figure 23 shows height profiles of EC obtained at local night in the northern polar cap (latitudes >70°) on December 8, 2006, when very intense fluxes of SEP electrons and low-energy protons are observed (see Figure 2). During that day we find that the SEP-associated maximum of EC at heights of ~80 km is comparable and even exceeds the maximum of nighttime *F*-layer at 1139 UT. Very high SEP ionization of lower ionosphere at altitudes of ~80 km causes absorption of the ionosonde radio-signals at altitudes below 100 km that resulted in the observed PCA effect.

Hence, we have found two independent evidences that the PCA observed on December 2006 are related to integral ionization effect produced mainly in the ionospheric *D* layer by the low energy protons with *E* >2.5 MeV and energetic electrons with *E* >100 KeV.

## 6. Conclusions

Using data acquired from five NOAA POES satellites, we perform a detail study of cutoff latitudes for protons and electrons during long-lasting SEP events on December 5 to 15, 2006.

Based on elliptical approach, we have investigated hourly variations of the cutoff boundaries for >300 keV electrons and protons with energies from 240 keV to >140 MeV.

We have developed a model of cutoff latitudes for protons and electrons in the northern and southern hemispheres. The cutoff latitude is represented as a function of rigidity, MLT, geomagnetic indices *Dst*, *Kp*, *AE* and dipole tilt angle *PS*.

The model is able to describe both tailward and duskward shifting of the cutoff boundaries. Such dynamics relates to variations of the geo-dipole tilt angle and intensification of the cross-tail current, field-aligned currents, and symmetrical and asymmetrical parts of the ring current.

Application of the model for prediction of PCA effects observed by a network of ionosondes at high-latitudes allowed us to reveal that the absorption of radio signal is mainly related to intense fluxes of >2.5 MeV protons with intensity of >100 $(cm^2\ s\ sr)^{-1}$ and >100 keV electrons with intensity of >2900 $(cm^2\ s\ sr)^{-1}$.

FORMOSAT-3/COSMIC measurements of electron content height profiles provide independent evidence that the PCA effects are caused by intense ionization of the ionospheric *D*-layer at altitudes of 75 to 85 km. That ionization is mainly produced by the >2.5 MeV protons and >100 keV electrons.

**Acknowledgements** The authors thank a team of NOAA's Polar Orbiting Environmental Satellites for providing experimental data about energetic electrons and protons, SPIDR NOAA (http://spidr.ngdc.noaa.gov/spidr/index.jsp) data resource for providing GOES data about solar X-ray fluxes, and Kyoto World Data Center for Geomagnetism (http://swdcwww.kugi.kyoto-u.ac.jp/index.html) for providing the *Dst* and *Kp* geomagnetic indices. We also acknowledge a team of SEP instrument suite on STEREO (http://www.srl.caltech.edu/STEREO/index.html) for providing data on relativistic electrons, and T. Von Rosenvinge at NASA/GSFC and CDAWeb for providing ion data from EPA instrument onboard Wind. We are grateful to David Evans for very valuable comments on POES/SEM instrument package. This work was supported by grants NSC-98-2111-M-008-002 and NSC-98-2111-M-008-019 from the National Science Council of Taiwan and by Ministry of Education under the Aim for Top University program at National Central University of Taiwan #985603-20. CADI operation is conducted in collaboration with the Canadian Space Agency.

**References**

Antonova, E. E., S. N. Kuznetsov, A. V. Suvorova (1989), Determination of some geomagnetic-field characteristics from the data of low-orbiting satellites, *Geomagn. Aeron.*, 29(4), 425-430.

Birch, M. J., J. K. Hargreaves, A. Senior, and B. J. I. Bromage (2005), Variations in cutoff latitude during selected solar energetic proton events, *J. Geophys. Res.*, 110, A07221, doi:10.1029/2004JA010833.

Blake, J. B., M. C. McNab, and J. E. Mazur (2001), Solar-proton polar-cap intensity structures as a test of magnetic field models, *Adv. Space Res.*, 28(12), 1753-1757.

Clilverd, M. A., C. J. Rodger, T. Moffat-Griffin, and P. T. Verronen (2007), Improved dynamic geomagnetic rigidity cutoff modeling: Testing predictive accuracy, *J. Geophys. Res.*, 112, A08302, doi:10.1029/2007JA012410.

Cummings, W. (1966), Asymmetric ring currents and the low-latitude disturbance daily variation, *J. Geophys. Res.*, 71(19), 4495-4503.

Dmitriev A., J.-K. Chao, D.-J. Wu, Comparative study of bow shock models using Wind and Geotail observations, *J. Geophys. Res.*, 108(A12), 1464, doi:10.1029/2003JA010027, 2003.

Dmitriev A.V., A.V. Suvorova, J.-K. Chao, Y.-H. Yang, Dawn-dusk asymmetry of geosynchronous magnetopause crossings, *J. Geophys. Res.*, 109, A05203 doi: 10.1029/2003JA010171, 2004.

Dmitriev, A. V., L.-C. Tsai, H.-C. Yeh, and C.-C. Chang (2008), COSMIC/FORMOSAT-3 tomography of SEP ionization in the polar cap, *Geophys. Res. Lett.*, 35, L22108, doi:10.1029/2008GL036146.

Droge, W. (2000), The rigidity dependence of solar particle scattering mean free paths, *The Astrophysical Journal*, 537, 1073-1079.

Evans, D. S., and M. S. Greer (2004), Polar Orbiting environmental satellite space environment monitor - 2 instrument descriptions and archive data documentation, NOAA technical Memorandum version 1.4, Space Environment Laboratory, Colorado.

Fanselow, J. L., and E. C. Stone (1972), Geomagnetic cutoff for cosmic-ray protons for seven energy intervals between 1.2 and 39 MeV, *J. Geophys. Res.*, 77(22) 3999-4009.

Frank L. A., J. A. Van Allen, and J. D. Graven ( 1964), Large diurnal variations of geomagnetically trapped and precipitated electrons observed at low altitudes, *J. Geophys. Res.*, 69, 3155-3167.

Galand, M. (2001), Introduction to special section: Proton precipitation into the atmosphere, J. Geophys. Res., 106(A1), 1-6.

Gotselyuk, Yu. V., S. N. Kuznetsov, I. Kimak, K. Kudela (1990), dependence of the polar cap structure on parameters of the interplanetary medium based on penetration deduced from solar cosmic ray electrons, *Ann. Geophys.*, 8(5), 369-376.

Gotselyuk, Yu. V., A. V. Dmitriev, S. N. Kuznetsov, A. V. Suvorova and N. Yu. Ganyushkina (1998), Dependence of polar cap size on interplanetary parameters according to "CORONAS-I" data, *Adv. Space Res.*, 22(9), 1323-1326.

Hajj, G. A., L. C. Lee, X. Pi, L. J. Romans, W. S. Schreiner, P. R. Straus, and C. Wang (2000), COSMIC GPS ionospheric sensing and space weather, *Terrestial, Atmospheric and Oceanic Sciences*, 11(1), 235-272.

Huston, S. L., and K. A. Pfitzer (1998), Space environment effects: low-altitude trapped radiation model, Marshall Space Flight Center, NASA/CR-1998-208593.

Ivanova, T. A., S. N. Kuznetsov, E. N. Sosnovets, L. V. Tverskaya (1985), Dynamics of low-latitude boundary for penetration of low-energy solar protons in the magnetosphere, *Geomagn. Aeron.*, XXV(1), 7-12 (in Russian).

Jayachandran, P. T., et al. (2009), Canadian High Arctic Ionospheric Network (CHAIN), *Radio Sci.*, 44, RS0A03, doi:10.1029/2008RS004046.

Kahler, S., and A. Ling (2002), Comparisons of high latitude E > 20MeV proton geomagnetic cutoff observations with predictions of the SEPTR model, *Ann. Geophys.*, 20, 997-1005.

Kress B. T., M. K. Hudson, K. L. Perry, P. L. Slocum (2004), Dynamic modeling of geomagnetic cutoff for the 23-24 November 2001 solar energetic particle event, Geophys. Res. Lett., 31, L04808, doi:10.1029/2003GL018599.

Liemohn, M., J. Kozyra, M. Thomsen, J. Roeder, G. Lu, J. Borovsky, and T. Cayton (2001), Dominant role of the asymmetric ring current in producing the stormtime Dst*, J. Geophys. Res., 106(A6), 10883-10904.

Leske, R. A., R. A. Mewaldt, E. C. Stone, and T. T. von Rosenvinge (2001), Observations of geomagnetic cutoff variations during solar energetic particle events and implications for the radiation environment at the Space Station, J. Geophys. Res., 106(A12), 30,011-30,022.

McDiarmid I. B., J. R. Burrows, and E. E. Budzinski (1975), Average characteristics of Magnetospheric Electrons (150 eV to 200 keV) at 1400 km, J.Geophys.Res., 80(1), 73-79.

Mewaldt R. A., C. M. S. Cohen, A. W. Labrador, R. A. Leske, G. M. Mason, M. I. Desai, M. D. Looper, J. E. Mazur, R. S. Selesnick, D. K. Haggerty (2005), Proton, helium, and electron spectra during the large solar particle events of October–November 2003, J. Geophys. Res., 110, A09S18, doi:10.1029/2005JA011038.

O'Brien, T. P., and R. L. McPherron (2002), Seasonal and diurnal variation of Dst dynamics, J. Geophys. Res., 107(A11), 1341, doi:10.1029/2002JA009435.

Ogliore, R. C., R. A. Mewaldt, R. A. Leske, E. C. Stone, and T. T. von Rosenvinge (2001), A direct measurement of the geomagnetic curoff for cosmic rays at space station latitudes, Proc. Int. Conf. Cosmic Rays, 27[th], 4112-4115.

Rodger C. J., M. A. Clilverd, P. T. Verronen, T. Ulich, M. J. Jarvis, E. Turunen (2006), Dynamic geomagnetic rigidity cutoff variations during a solar proton event, J. Geophys. Res., 111, A04222, doi:10.1029/2005JA011395.

Sergeev, V. A., S. N. Kuznetsov, Yu. V. Gotselyuk (1987), Dynamics of high-latitude magnetosphere structure derived from data on solar electrons, *Geomagn. Aeron.*, XXVII(3), 440-447 (in Russian).

Shea, M.A., and Smart, D.F. (2004), Preliminary study of cosmic rays, geomagnetic field changes and possible climate changes. *Adv. Space Res.*, 34, 420–425.

Smart, D.F., and M.A. Shea (1994), Geomagnetic cutoffs: A review for space dosimetry applications, *Adv. Space Res.*, 14(10), 787-796.

Smart, D.F., and M.A. Shea (2001), A comparison of the Tsyganenko model predicted and measured geomagnetic cutoff latitudes, *Adv. Space Res.*, 28(12), 1733-1738.

Smart, D. F., and M. A. Shea (2003), The space-developed dynamic vertical cutoff rigidity model and its applicability to aircraft radiation dose, *Adv. Space Res.*, 32(1), 103-108.

Smart, D. F., M. A. Shea, and E. O. Fluckiger (2000), Magnetospheric models and trajectory computations, *Space Sci. Rev.*, 93, 281-308.

Smart, D.F., M.A. Shea, A.J. Tylka, P.R. Boberg (2006), A geomagnetic cutoff rigidity interpolation tool: Accuracy verification and application to space weather, *Adv. Space Res.*, 37, 1206-1217.

Williams D. J., G. D. Mead (1965), Nightside Magnetosphere Configuration as Obtained from Trapped Electrons at 1100 km, *J.Ceophys. Res.*, 70, 3017-3024.

Zhang, M., R. B. McKibben, C. Lopate, J. R. Jokipii, J. Giacalone, M.-B. Kallenrode, and H. K. Rassoul (2003), Ulysses observations of solar energetic particles from the 14 July 2000 event at high heliographic latitudes, J. Geophys. Res., 108, 1154, doi:10.1029/2002JA009531.

**Table 1.** Number of cutoff latitudes determined for various SEP species

| SEP species | Rigidity, MV | Northern PC | Southern PC |
|---|---|---|---|
| e >300 keV | 0.63 | 417 | 426 |
| p 240 – 800 keV | 22 | 501 | 500 |
| p 0.8 – 2.5 MeV | 40 | 588 | 615 |
| p 2.5 – 6.9 MeV | 70 | 573 | 609 |
| p >6.9 MeV | 117 | 381 | 314 |
| p >16 MeV | 180 | 1127 | 1084 |
| p >36 MeV | 270 | 843 | 839 |
| p >70 MeV | 380 | 553 | 558 |
| p >140 MeV | 600 | 202 | 209 |

**Table 2.** Best-fit coefficients obtained for the semi-axes of cutoff boundaries during magnetic quiet

| $\sin^4(as_q)$ | $A$, $10^{-2}$ | $B$, $10^{-5}$ MV$^{-1}$ | $C$, $10^{-2}$ | $D$, $10^{-2}$ MV$^{-1}$ |
|---|---|---|---|---|
| Northern PC | 2.41 | 6.23 | -1.26 | -2.19 |
| Southern PC | 2.46 | 6.63 | -.922 | -1.28 |
| $\sin^4(bs_q)$ | | | | |
| Northern PC | 2.11 | 6.21 | -1.19 | -1.81 |
| Southern PC | 1.95 | 7.02 | -.895 | -.817 |

**Table 3.** Coefficients for elliptical parameters of the cutoff boundaries for protons

| Northern PC | $A_0$ | $B_0$ | $A_{Dst}\cdot10^{-2}$ | $B_{Dst}\cdot10^{-2}$ | $A_{Kp}\cdot10^{-2}$ | $B_{Kp}\cdot10^{-3}$ | $A_{PS}\cdot10^{-2}$ | $B_{PS}\cdot10^{-2}$ |
|---|---|---|---|---|---|---|---|---|
| $\Delta as$ | -.568 | 0. | -9.74 | 1.16 | 8.55 | -7.46 | | |
| $\Delta bs$ | -.648 | 0. | -9.54 | 1.16 | 7.66 | -7.49 | | |
| $Xo$ | -0.29 | 0.13 | 3.68 | -.544 | -2.94 | 3.40 | | |
| $Yo$ | -9.26 | 1.53 | -5.43 | .827 | -6.60 | 10.5 | -7.40 | 1.12 |
| $\varphi$ | -54.9 | 11.3 | | | | | | |
| Southern PC | | | | | | | | |
| $\Delta as$ | -0.740 | 0. | -6.40 | 0.597 | 11.5 | -13.7 | | |
| $\Delta bs$ | -0.400 | 0. | -5.14 | 0.445 | 5.21 | -3.51 | | |
| $Xo$ | -0.302 | 0. | 5.53 | -0.729 | -4.11 | 5.51 | | |
| $Yo$ | -8.54 | 1.43 | -3.93 | .433 | -7.89 | 11.7 | -7.27 | 1.30 |
| $\varphi$ | -44.5 | 7.83 | | | | | | |

**Table 4.** Coefficients for elliptical parameters of the cutoff boundaries for electrons

| Northern PC | $P_0$ | $P_{Dst}\cdot10^{-2}$ | $P_{Kp}\cdot10^{-2}$ | $P_{PS}\cdot10^{-2}$ | $P_{AE}\cdot10^{-4}$ |
|---|---|---|---|---|---|
| *as* | 18.9 | -4.3 | 9.86 | | |
| *bs* | 17.1 | -4.97 | 8.42 | | |
| *Xo* | 0.758 | .482 | -2.58 | | |
| *Yo* | -5.47 | | | -4.68 | 4.67 |
| $\varphi$ | 17.6 | -67. | | 178. | |
| Southern PC | | | | | |
| *as* | 18.5 | -2.83 | 12.1 | | |
| *bs* | 17.0 | -1.89 | 8.87 | | |
| *Xo* | 0.506 | 2.73 | -2.55 | | |
| *Yo* | -5.6 | | | -4.02 | 3.59 |
| $\varphi$ | -31.2 | 22.5 | | -78.8 | |

**Table 5.** Selected CADI stations.

| Code | Station | Lat | Lon | mLat | mLon | $\Lambda_O$ |
|---|---|---|---|---|---|---|
| EUR | Eureka | 80.0 | 274.0 | 86.70 | 237.2 | 5.35 |
| R_B | Resolute Bay | 74.8 | 265.0 | 83.15 | 290.5 | 7.21 |
| RAN | Rankin | 62.8 | 267.9 | 72.84 | 324.0 | 17.64 |

**Table 6.** Logic table

| PCA observed | $\Lambda_m\geq\Lambda_O$ | $\Lambda_m<\Lambda_O$ |
|---|---|---|
| Yes | A | B |
| No | C | D |

**Table 7.** Main statistical numbers for PCA prediction

| | Threshold (cm$^2$ s sr)$^{-1}$ | PCP | OUR |
|---|---|---|---|
| **p >2.5 MeV*** | **100** | **.75** | **.52** |
| Rankin | 200 | .79 | .49 |
| Resolute Bay | 85 | .75 | .49 |
| Rankin & Resolute Bay | 130 | .76 | .53 |
| **p >2.5 MeV & e >300 keV*** | **100 & 2900** | **.77** | **.50** |
| Rankin | 300 & 2900 | .80 | .51 |
| Resolute Bay | 100 & 2900 | .75 | .49 |
| Rankin & Resolute Bay | 150 & 2900 | .77 | .50 |

* For all stations: Rankin, Resolute Bay and Eureka

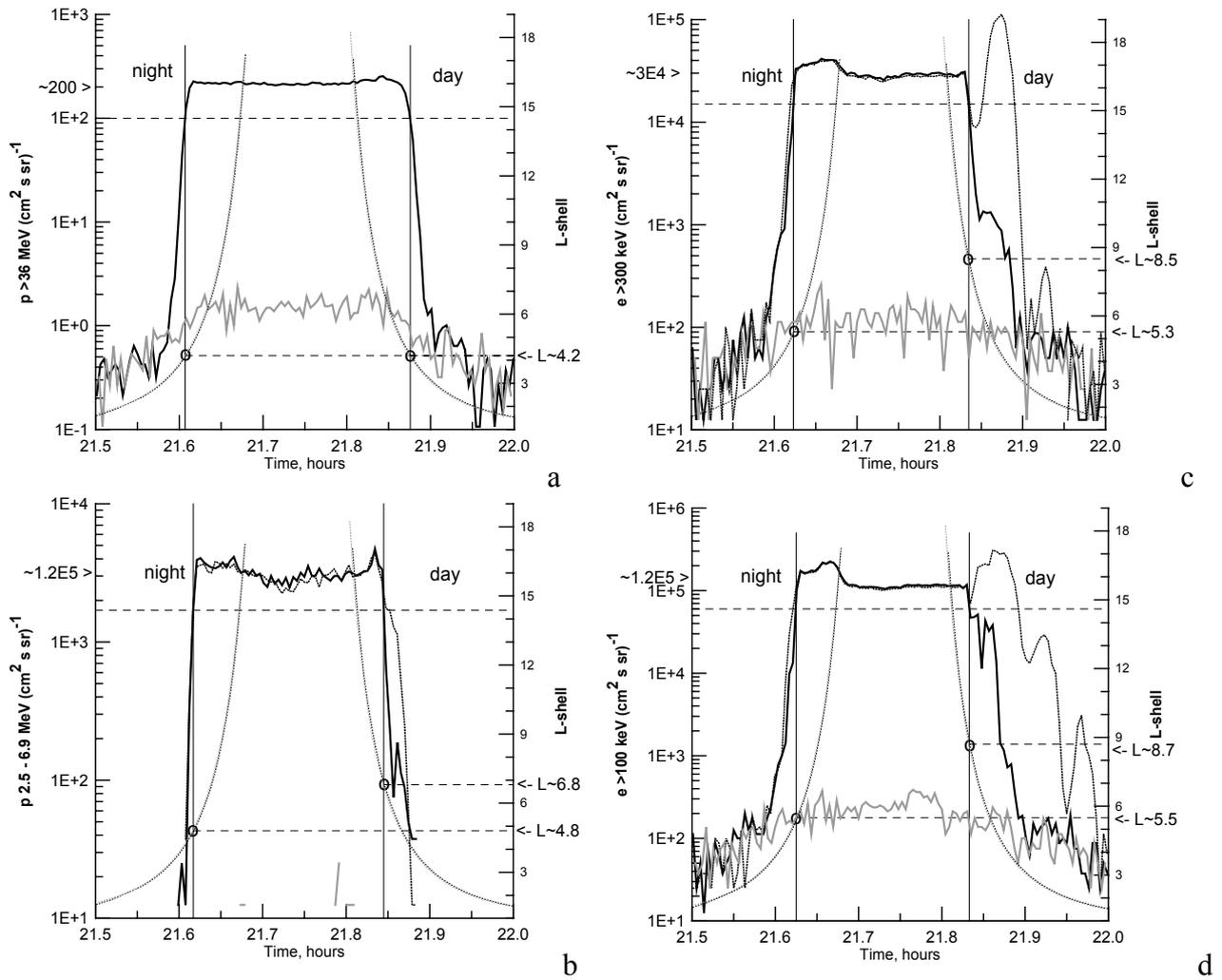

Figure 1. Fluxes of energetic particles observed by P17 in the northern polar cap during quiet day on December 5 (gray thick curves) and during SEP event on December 7, 2006 (black thick curves): (a) >36 MeV protons (omni-directional), (b) 2.6 − 6.9 MeV protons, (c) >300 keV electrons, and (d) >100 keV electrons. Fluxes of particles with pitch angles of 0° and 90° are shown by solid and dotted thick curves, respectively. *L*-shell (right axis) is presented by thin gray and black dotted lines, respectively, for quiet day and for SEP event. Vertical thin lines depict time moments when the SEP flux is equal to 50% of the mean SEP intensity in the polar cap (shown by horizontal dashed lines). Horizontal dashed segments indicate the *L*-shells for those moments.

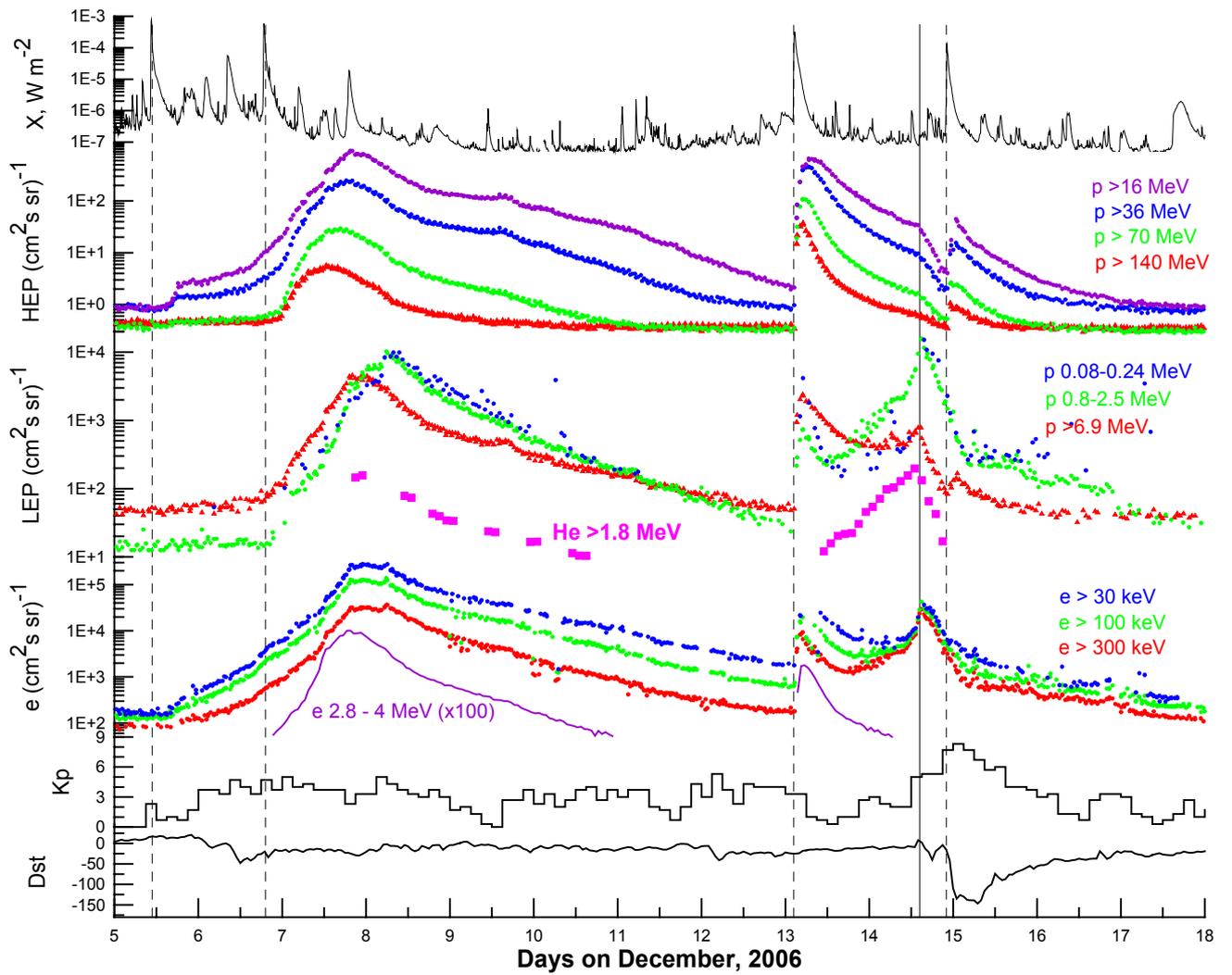

Figure 2. SEP events on December 2006 (from top to bottom): solar 1-8 Å X-ray flux observed by GOES-11; POES observations of HEP, LEP and energetic electrons in the northern polar cap, *Kp* and *Dst* geomagnetic indices. Violet squares show hourly averaged fluxes of >1.8 MeV helium measured by the SOHO/ERNE instrument. Relativistic electrons with $E = 2.8 - 4$ MeV (violet curve) are measured by HET/STEREO instrument. Vertical dashed lines indicate X-class solar flares, and vertical solid line indicates IS arrival to the Earth.

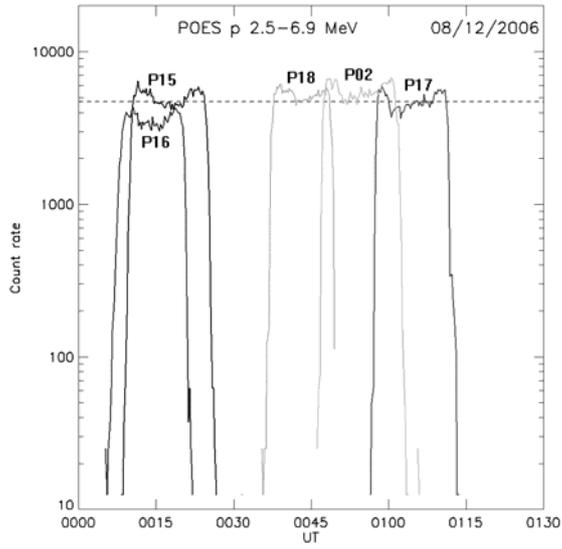

a

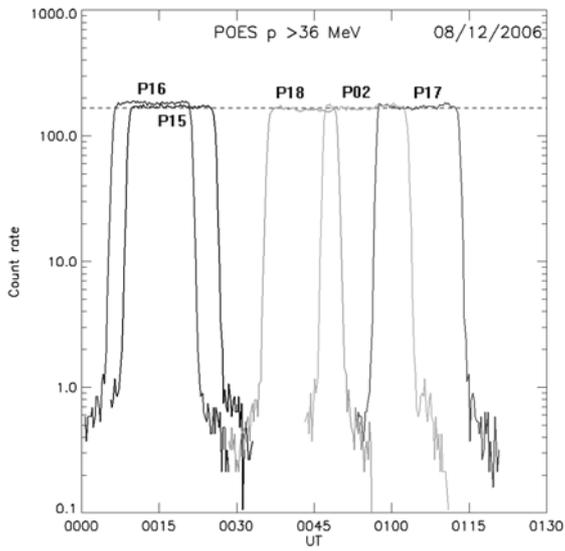

b

Figure 3. Fluxes of (a) 2.5 – 6.9 MeV and (b) >36 MeV proton measured in the northern polar cap by different POES satellites. The P16 satellite reveals ~20% underestimation and ~5% overestimation of the LEP and HEP fluxes, respectively.

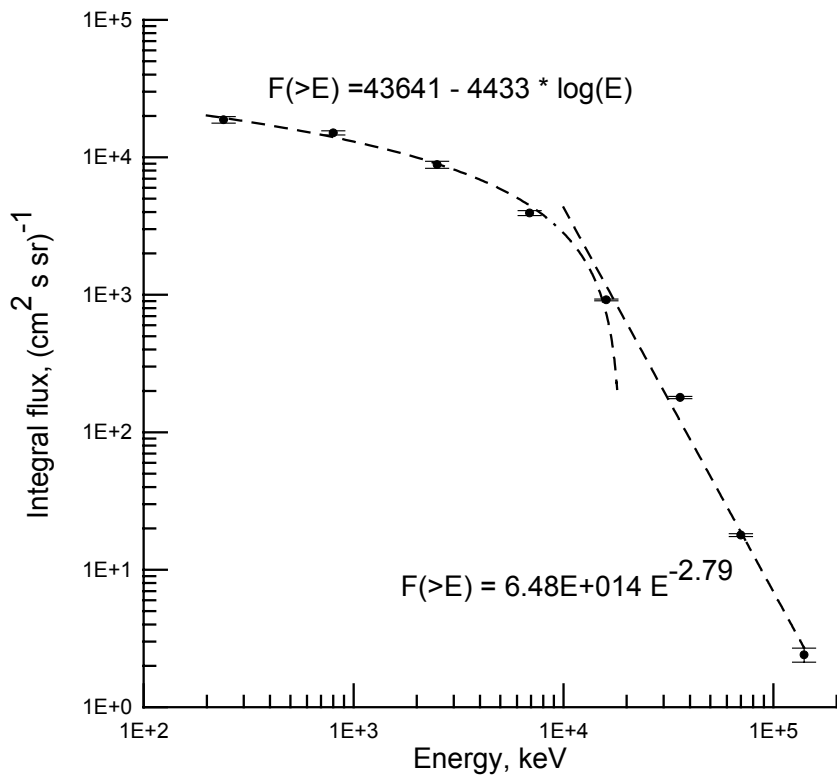

Figure 4. Average integral spectrum of solar protons detected by POES satellites in the polar cap at 0000 – 0100 UT on December 8, 2006. The LEP fluxes with energies below ~16 MeV can be fitted by a logarithm function, while the HEP with energies above ~16 MeV are well described by a power spectrum.

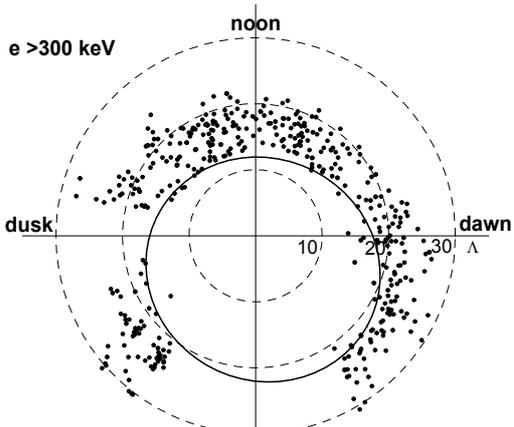

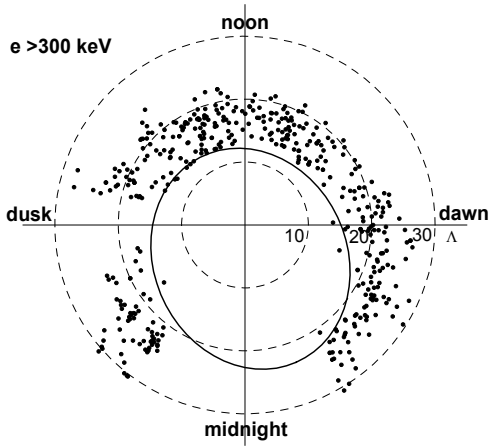

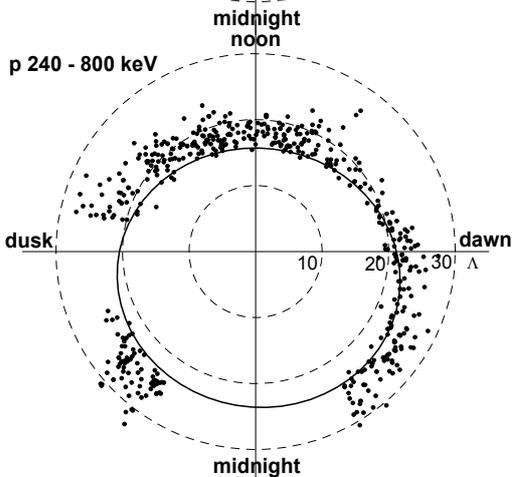

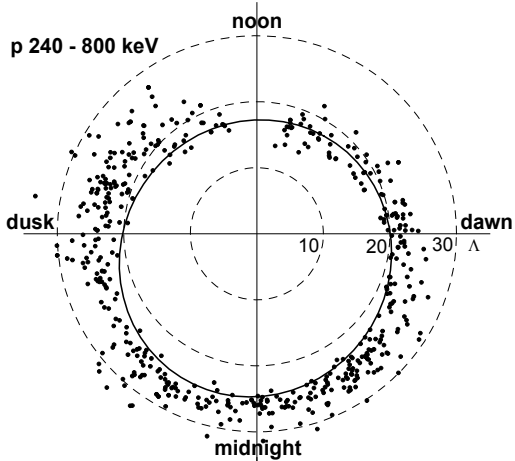

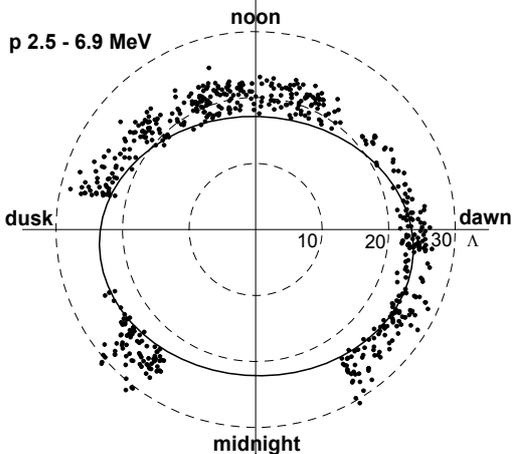

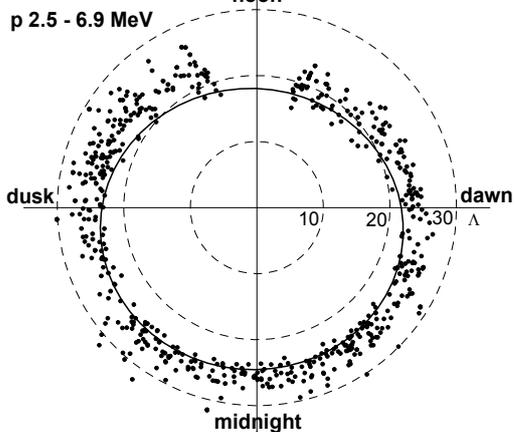

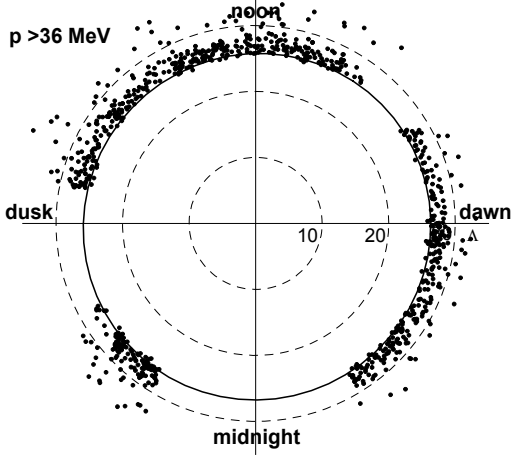

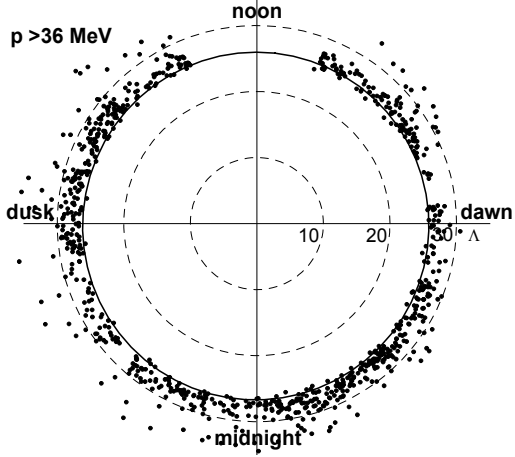

Figure 5. Figure 5. Scatter plots of cutoff latitudes (Λ vs MLT) determined for various SEP species in the northern (left) and nouthern (right) hemispheres during December 2006. Solid ellipses indicate best fits for magnetic quiet at ~1200 UT on December 9, 2006.

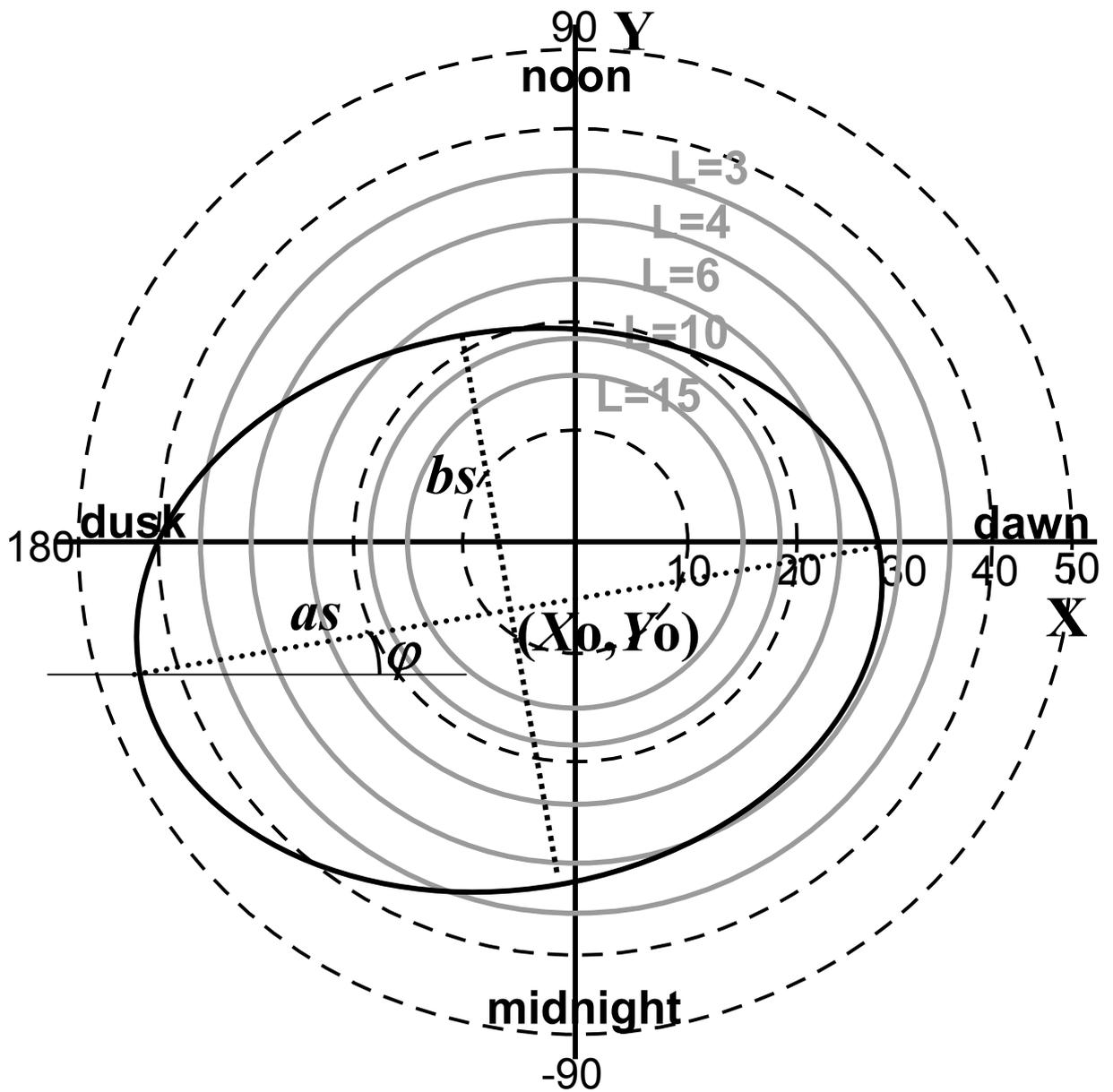

Figure 6. Coordinate system {rMLT, Λ} and elliptical fitting parameters. Here Λ is invariant co-latitude, rMLT is rotated magnetic local time (in degrees) calculated from dawn (X-axis). Gray circles indicate various L-shells. The ellipse is defined by 5 parameters: major *as* and minor *bs* semiaxes, location of the center (*X*o,*Y*o) and azimuth angle *φ* of the major semiaxis.

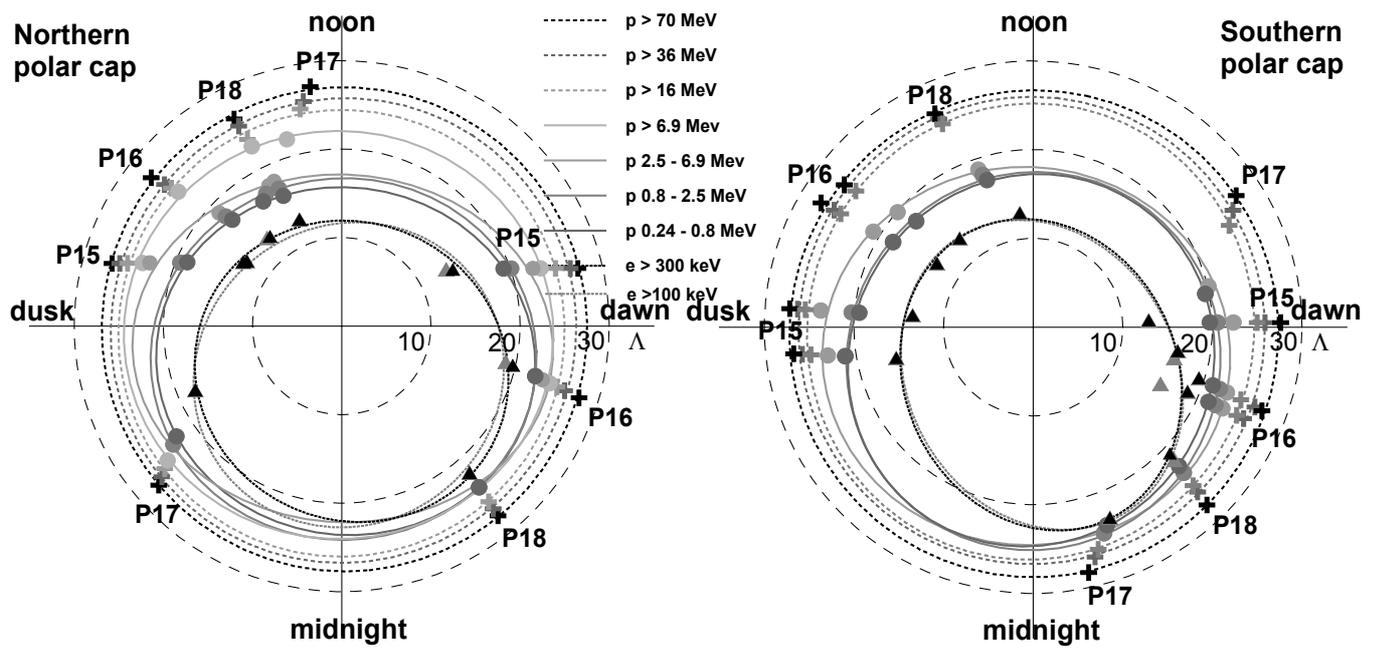

Figure 7. Fitting the SEP cutoff boundaries during magnetic quiet in the northern (left) and southern (right) hemispheres. Cutoff latitudes determined for the HEP, LEP and electrons are indicated, respectively, by crosses, circles, and triangles. The passes are indicated by corresponding satellite abbreviations.

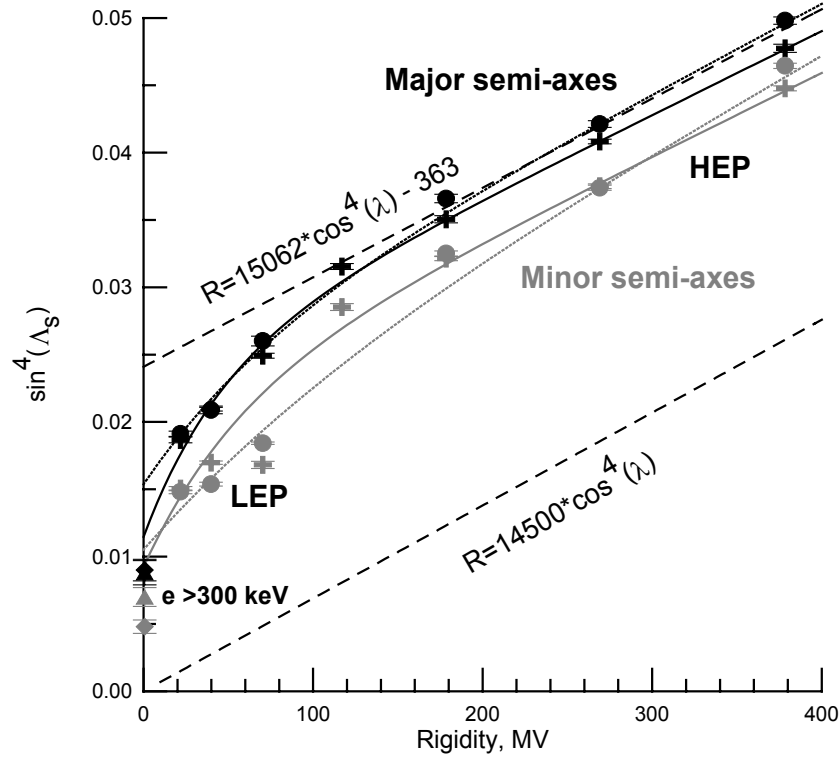

Figure 8. Major (black symbols) and minor (gray symbols) semiaxes of the cutoff boundaries for protons and >300 keV electrons, respectively, in the northern (crosses and triangles) and southern (circles and diamonds) hemispheres during magnetic quiet. Approximations of the major (minor) semiaxis for protons are shown by black (gray) solid and dotted curves, respectively, for the northern and southern hemisphere. Strait dashed lines are predictions by *Smart and Shea* [1994] (lower) and by *Ogliore et al.* [2001] (upper).

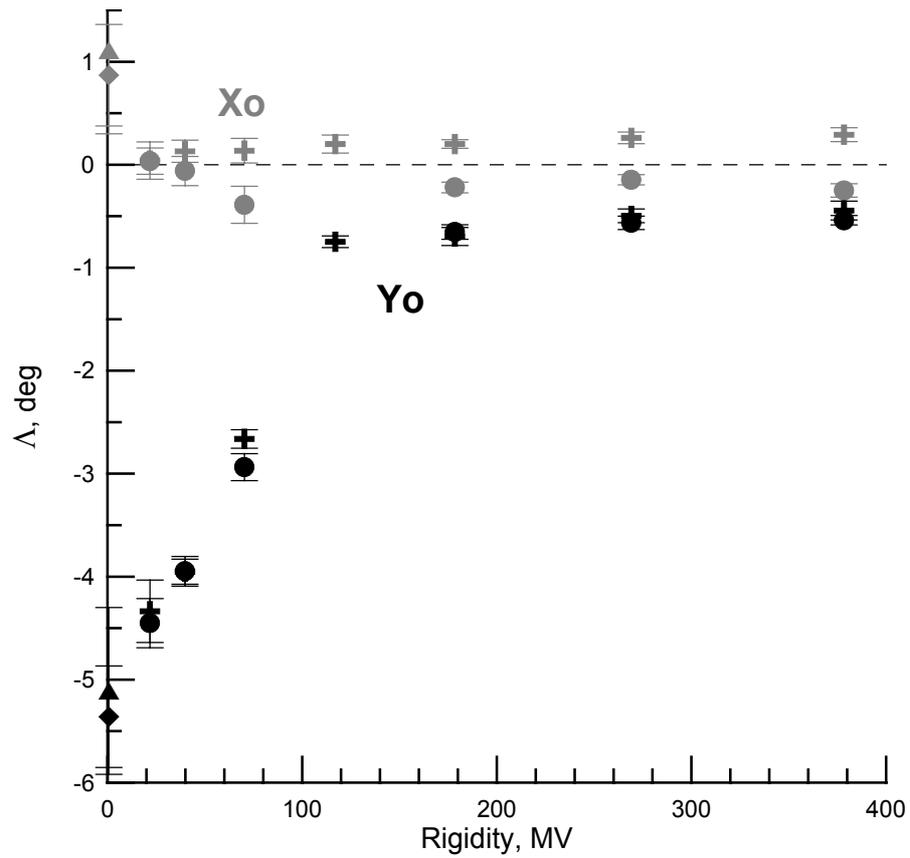

Figure 9. Coordinates *X*o (gray symbols) and *Y*o (black symbols) of the cutoff boundary centers determined from elliptical fitting for protons and >300 keV electrons, respectively, in the northern (crosses and triangles) and southern (circles and diamonds) hemisphere during magnetic quiet period.

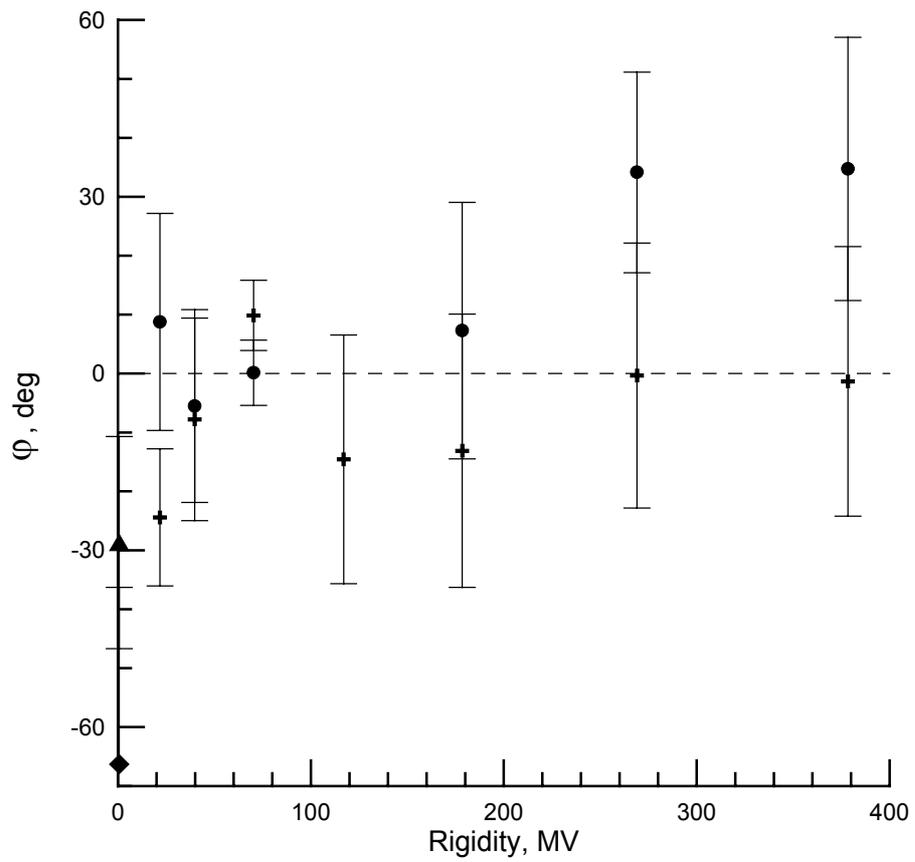

Figure 10. Scatter plot of azimuth angles versus rigidity for protons and > 300 keV electrons, respectively, in the northern (crosses and triangle) and southern (circles and diamond) hemispheres during magnetic quiet.

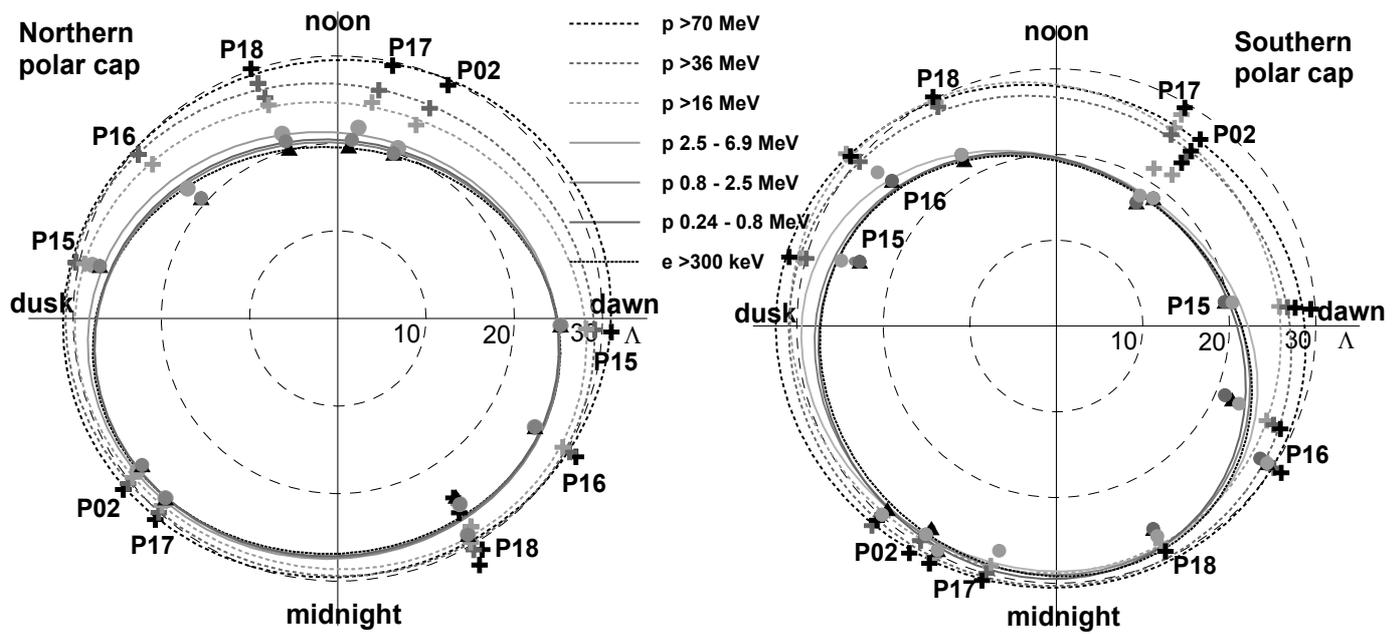

Figure 11. SEP cutoff boundaries during magnetic storm at 1700 UT on December 14, 2006 in the northern (left) and southern (right) hemispheres. The cutoff latitudes determined for HEP, LEP and electrons are indicated by crosses, circles, and triangles, respectively. The passes are indicated by corresponding satellite abbreviations.

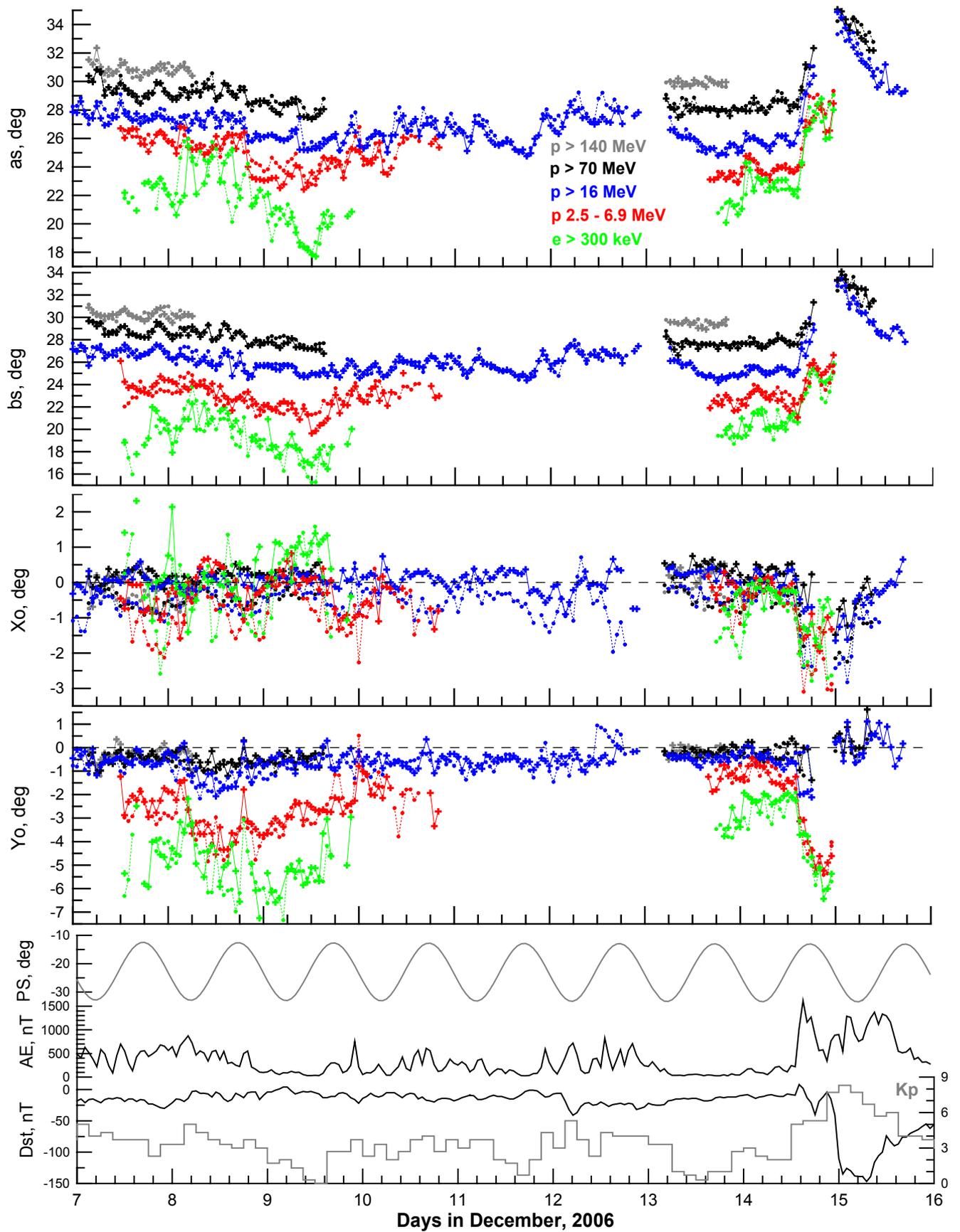

Figure 12. Variations of the fitting parameters for cutoff boundaries during 7 to 16 December, 2006 (from top to bottom): major *as*, and minor *bs* semiaxes, coordinates of center *X*o, and *Y*o, geodipole tile angle *PS*, *AE* index, *Dst* and *Kp* (gray histogram, right axis) indices of geomagnetic activity. The parameters derived in the northern and southern hemisphere are depicted by solid curves with crosses and doted curves with circles, respectively.

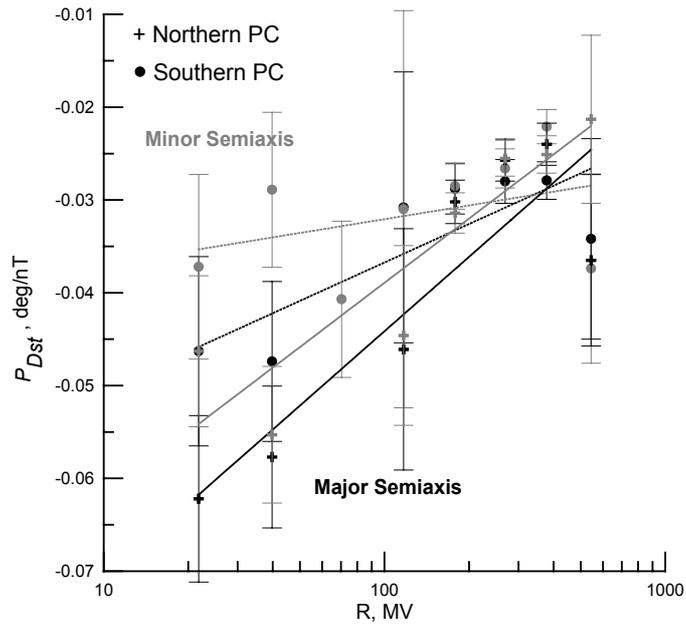

a

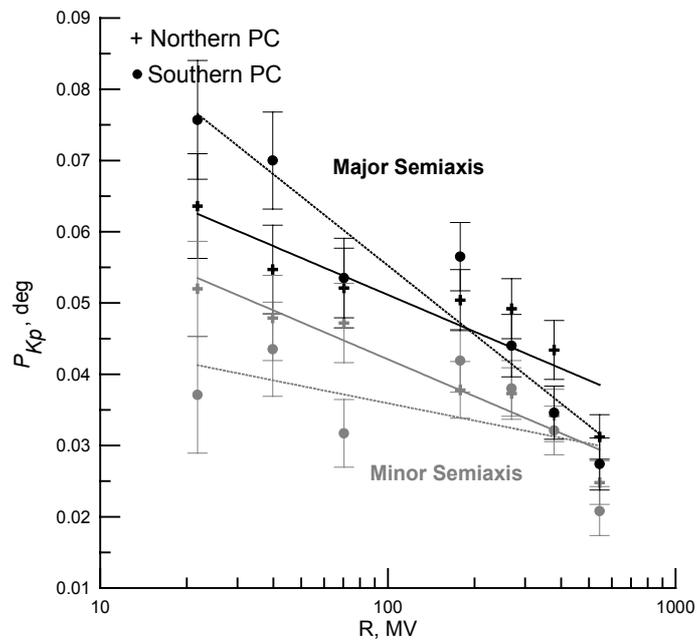

b

Figure 13. Dependence on rigidity for regression coefficients (a) $P_{Dst}$ and (b) $P_{Kp}$ obtained for major (black symbols) and minor (gray symbols) semiaxes in the northern (crosses) and southern (circles) hemisphere. Linear fittings in the logarithmic scale for rigidity are depicted by black (gray) solid and dotted strait lines for major (minor) semiaxes in the northern and southern hemisphere, respectively.

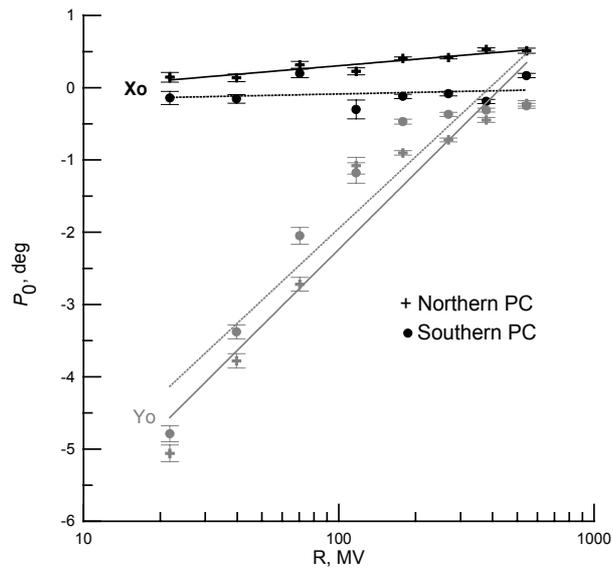

a

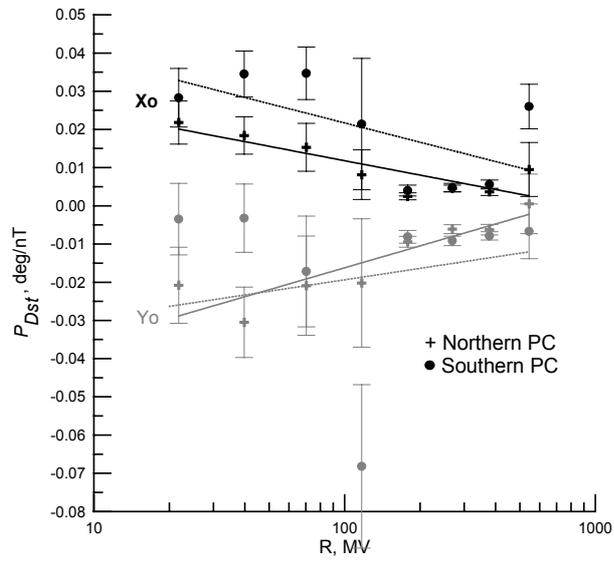

b

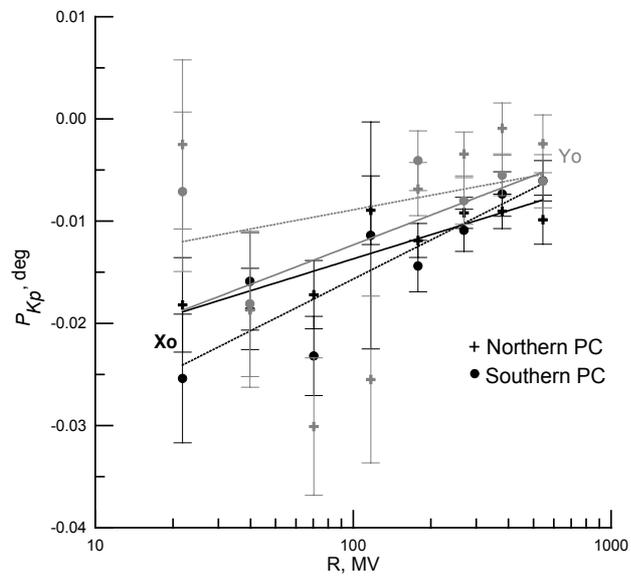

c

Figure 14. Dependence on rigidity for regression coefficients (a) $P_0$, (b) $P_{Dst}$ and (c) $P_{Kp}$ obtained for coordinates $X$o (black symbols) and $Y$o (gray symbols) in the northern (crosses) and southern (circles) hemisphere. Linear fittings in the logarithmic scale for rigidity are depicted by black (gray) solid and dotted strait lines for $X$o ($Y$o) in the northern and southern hemisphere, respectively.

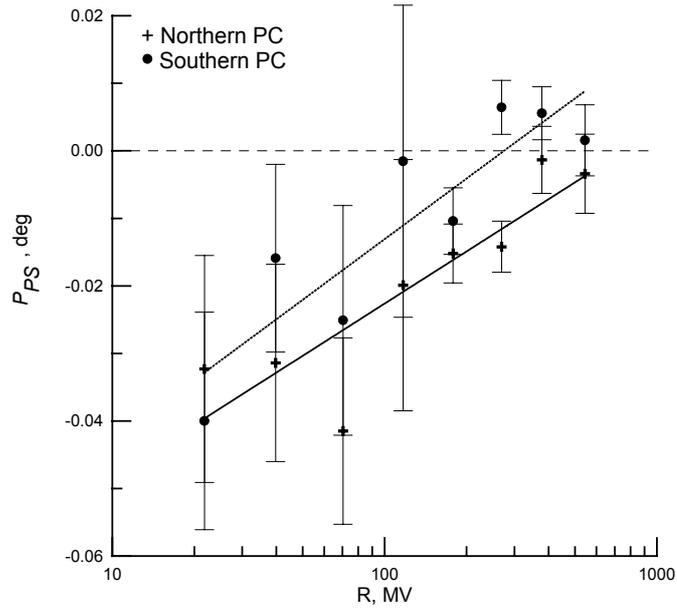

Figure 15. Regression coefficients $P_{PS}$ obtained for coordinates $Y_o$ versus proton rigidity in the northern (crosses) and southern (circles) hemisphere. Linear fittings in the logarithmic scale for rigidity are depicted by solid (dotted) strait lines in the northern (southern) hemisphere.

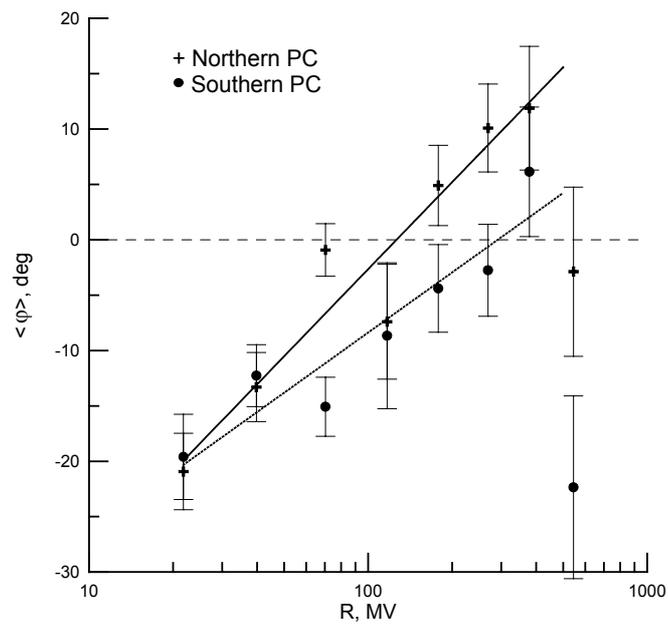

Figure 16. Mean azimuth angle versus proton rigidity in the northern (crosses) and southern (circles) hemisphere. Linear fittings in the logarithmic scale for rigidity are depicted by solid (dotted) strait lines in the northern (southern) hemisphere.

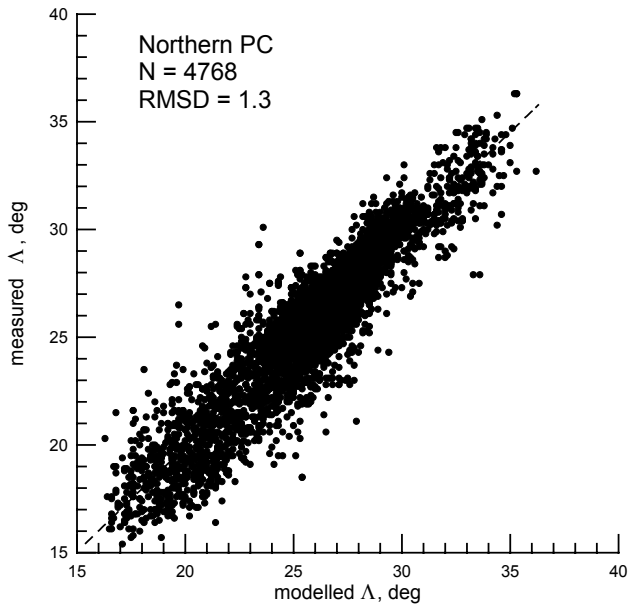

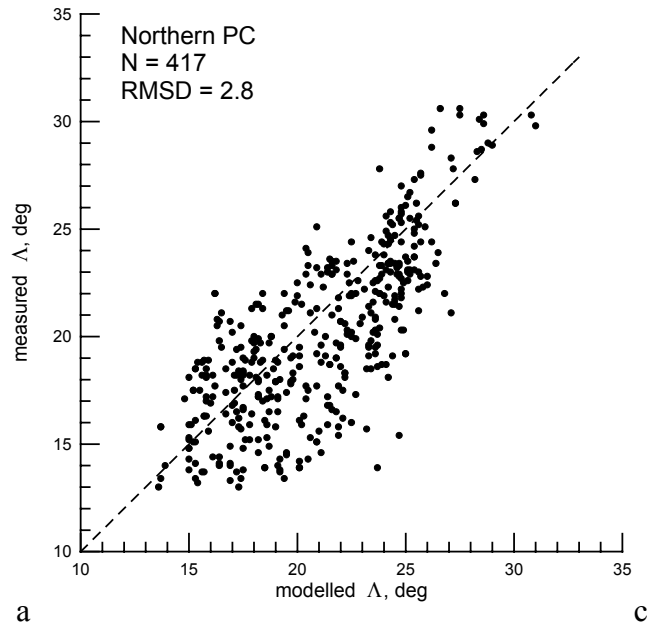

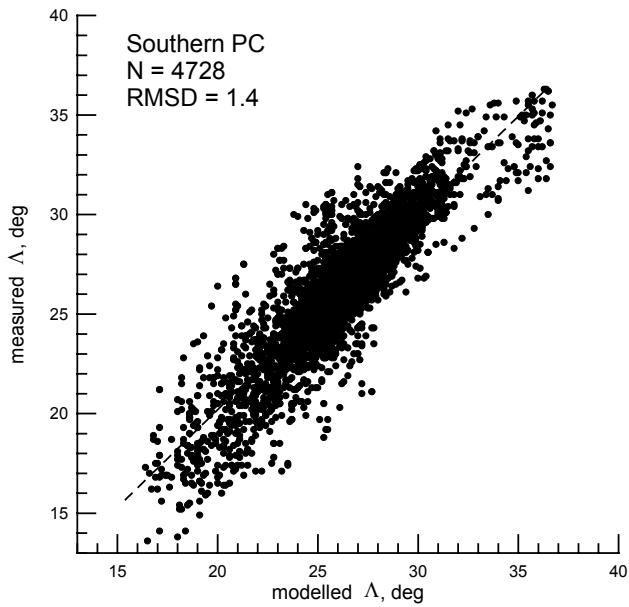

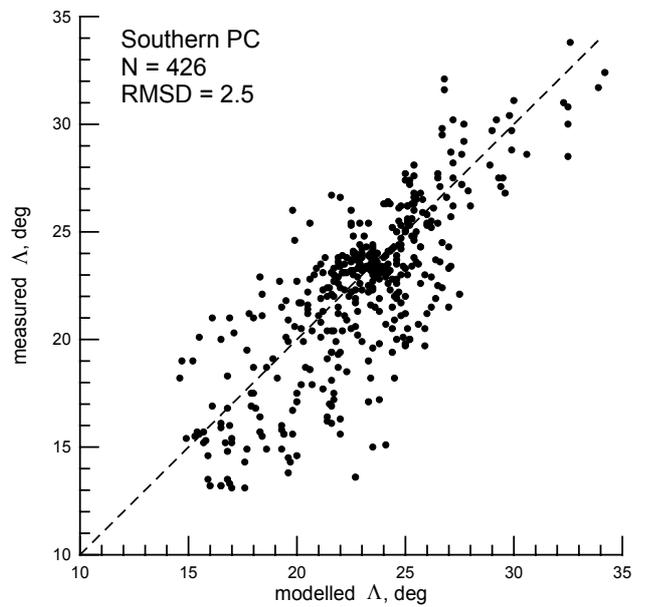

Figure 17. Comparison of determined cutoff latitudes with the model predictions for protons (a and b) and electrons (c and d), respectively, in the northern (a and c) and southern (b and d) hemisphere.

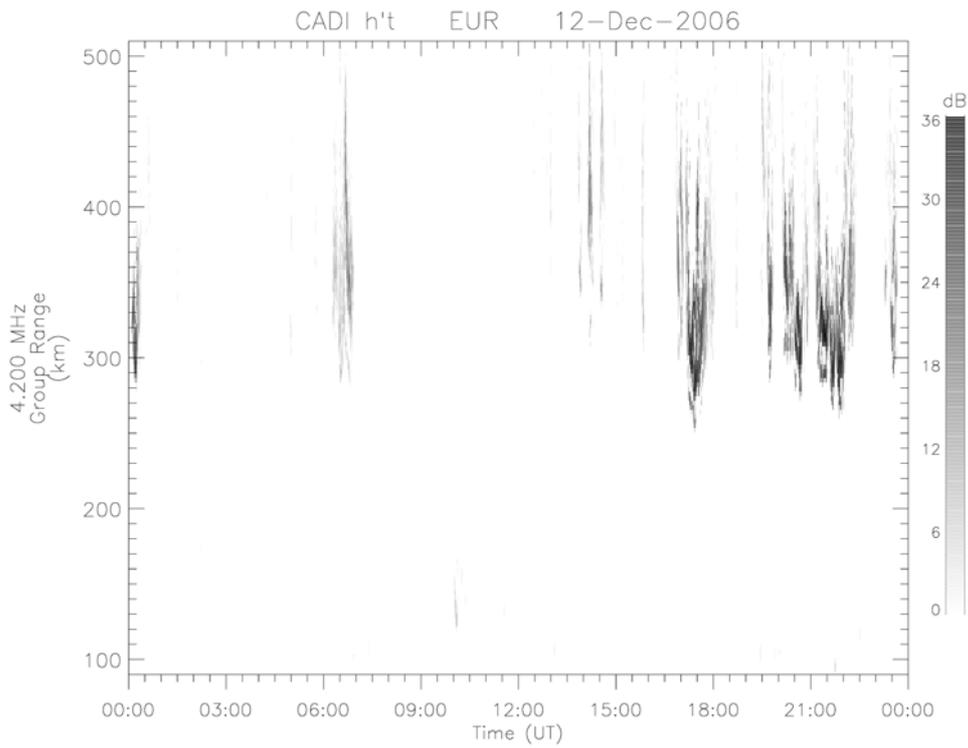

a

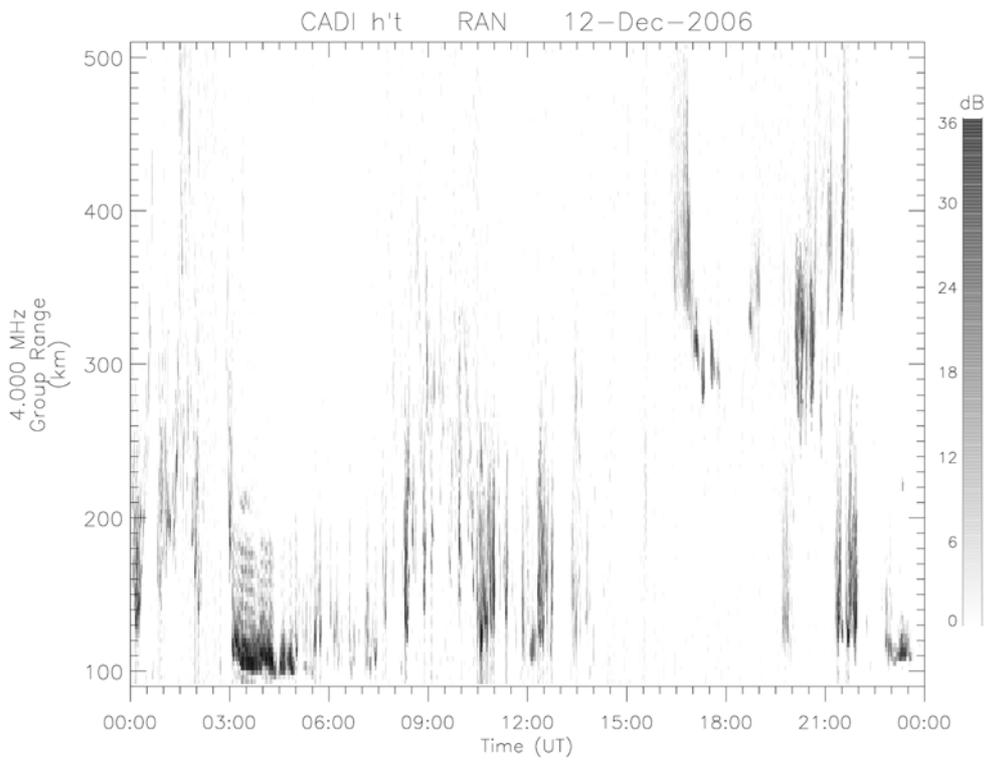

b

Figure 18. Dynamics of a group range of ionosonde radio signal at a given frequency of ionogram observed at (a) Eureka and (b) Rankin stations on December 12, 2006. The PCA effect is observed at Eureka at 0100 – 0600 UT and 0700 – 1400 UT, when very weak reflected radio signal is detected at altitudes above 100 km.

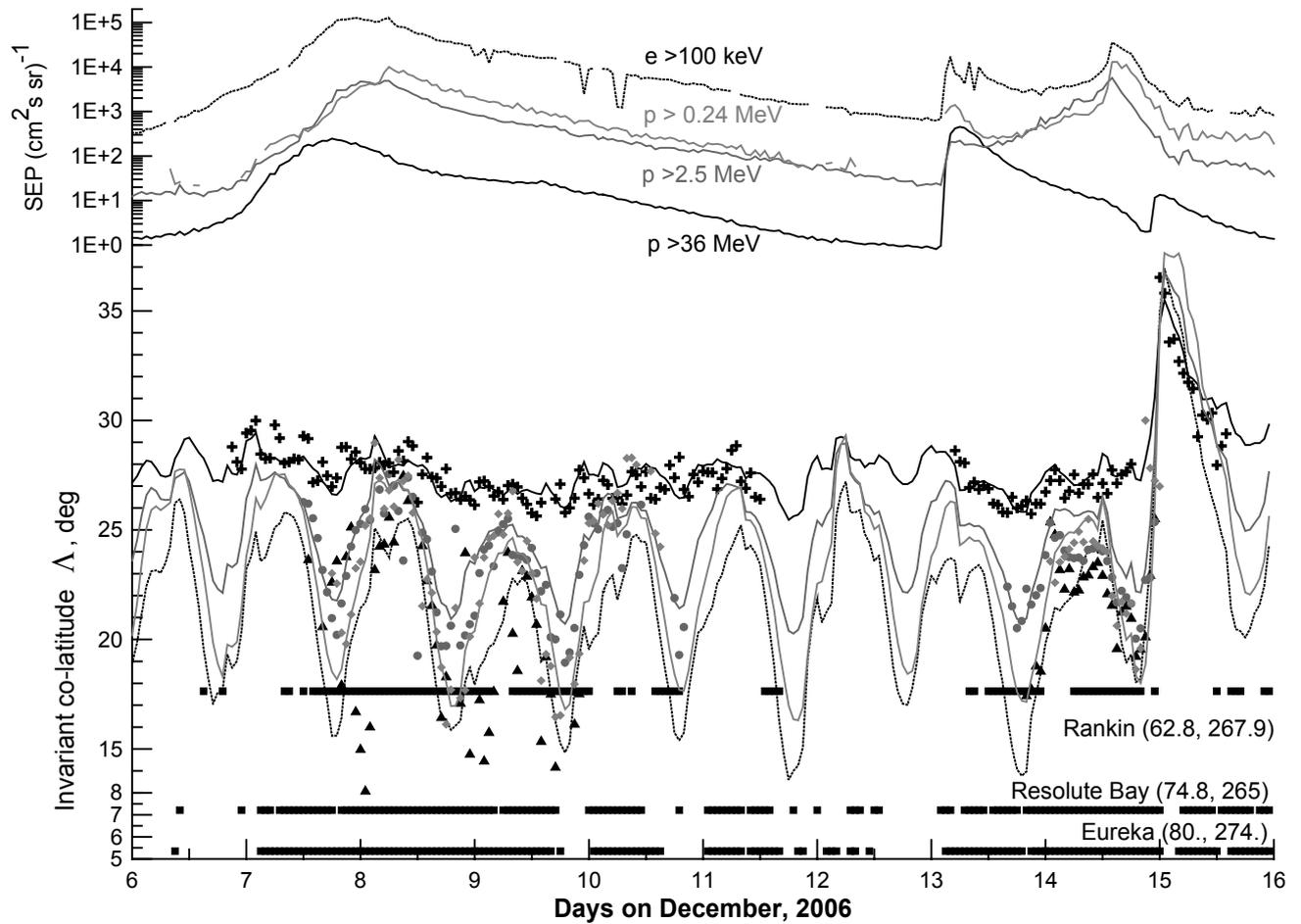

Figure 19. SEP fluxes and polar cap absorption effects observed in December 2006: (top panel) fluxes of >300 keV electrons (black dotted curve), and protons with energies >240 keV (light gray curve), >2.5 MeV (gray curve) and >36 MeV (black solid curve); (botton panel) PCA observed at stations Rankin, Resolute Bay and Eureka of CADI network (black squares) and model prediction together with elliptical approach of cutoff invariant co-latitudes at MLT of Rankin for the >300 keV electrons (black dotted curve and triangles), and protons with energies >240 keV (light gray curve and diamonds), >2.5 MeV (gray curve and circles) and >36 MeV (black curve and crosses). See details in the text.

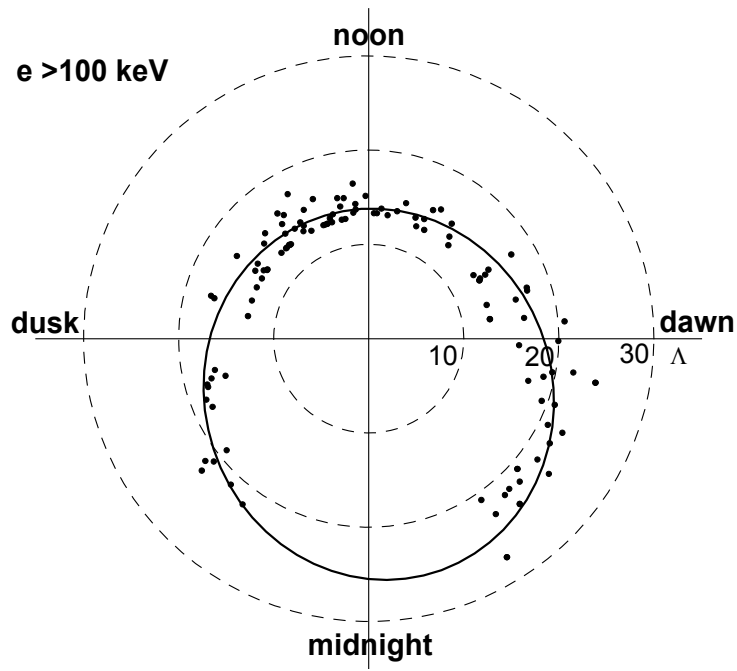

Figure 20. Scatter plots of cutoff latitudes determined for >100 keV electrons in coordinate system {MLT, Λ} during low substorm activity. Solid ellipse indicates the cutoff boundary obtained as a best fit of the latitudes.

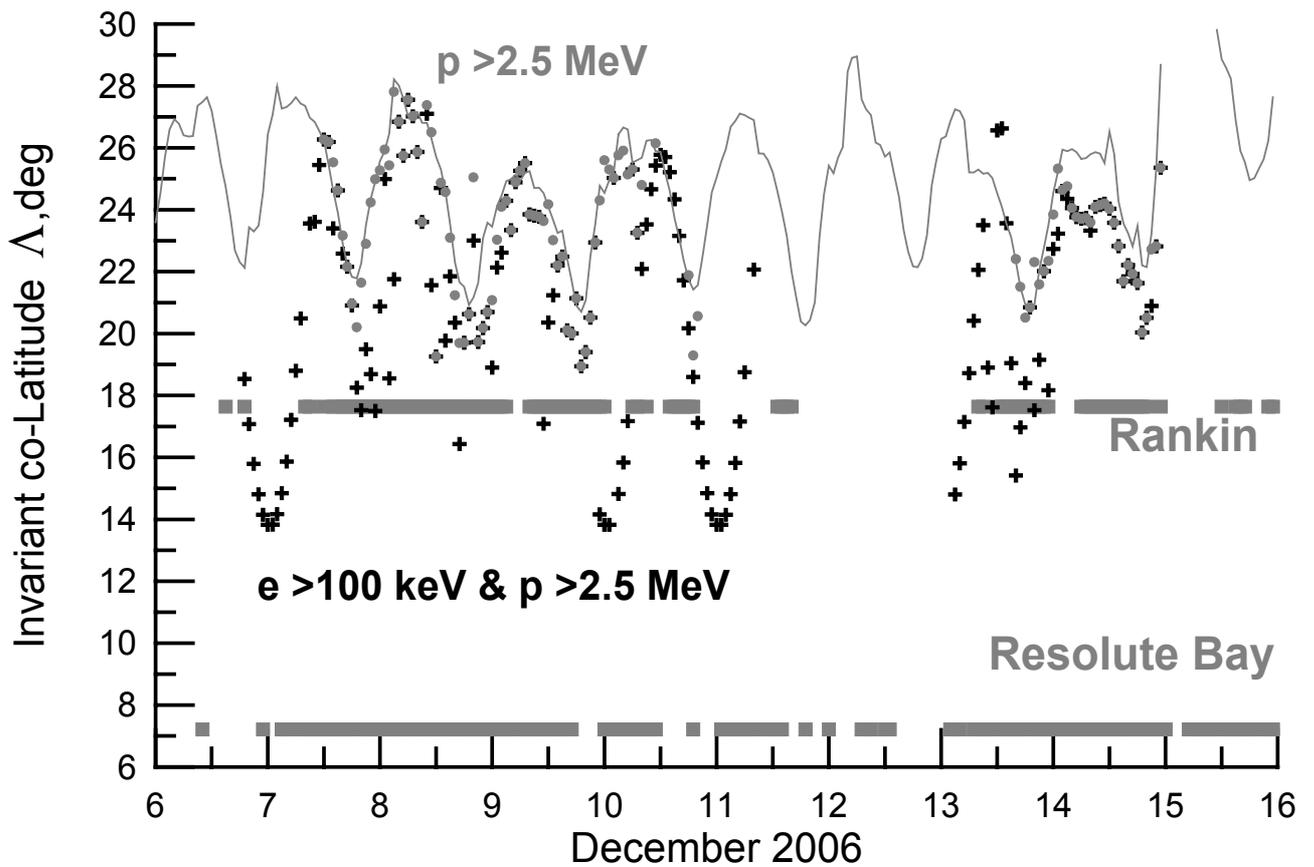

Figure 21. Comparison of PCA (gray squares) observations at Rankin and Resolute Bay with the cutoff boundary predicted for the SEP at MLT of Rankin. Gray curve indicates the model prediction of cutoff latitude for the >2.5 MeV protons. Gray circles indicates the elliptical approach of cutoff latitude for the >2.5 MeV protons, when those fluxes exceed 200 $(cm^2 s sr)^{-1}$. Crosses depict the minimum cutoff latitude accessible for the >2.5 MeV protons or >100 keV electrons, which fluxes exceed the threshold value of 2900 $(cm^2 s sr)^{-1}$. See details in the text.

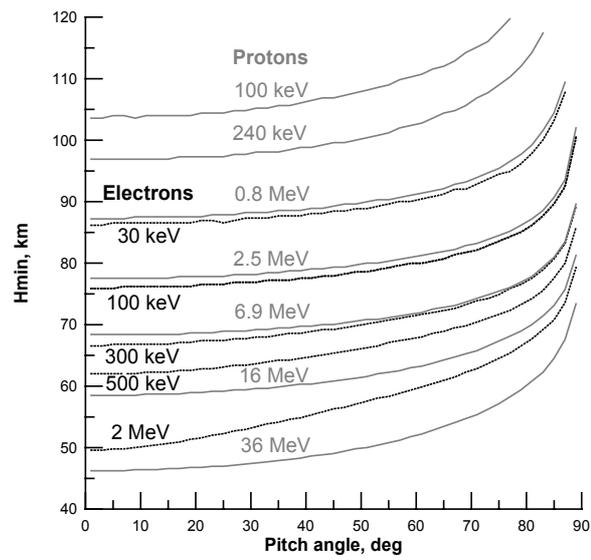

Figure 22. Minimal accessible heights for SEP electrons and protons of various energies versus incident pitch angle.

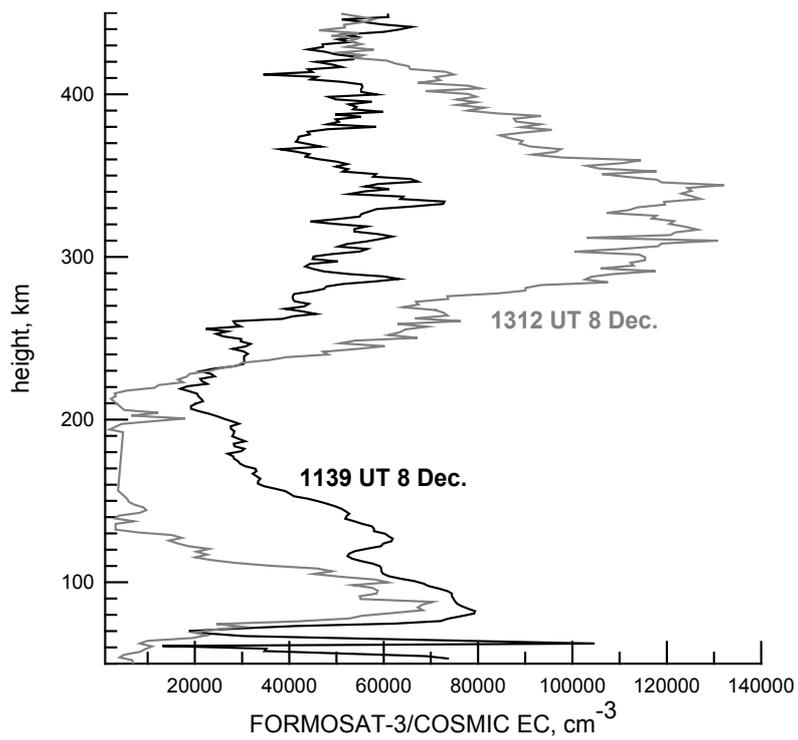

Figure 23. Height profiles of EC derived from FORMOSAT-3/COSMIC radio occultation measurements at local night in the Northern polar cap on December 8, 2006. The SEP-associated maximum of EC at low heights of ~80 km is comparable and even exceeds the maximum of nighttime F-layer.